\documentclass[preprint]{elsarticle}

\usepackage{graphicx}
\usepackage{epsfig}
\usepackage{amssymb}

\journal{Nuclear Physics A}

\newcommand{\ba}{\begin{eqnarray}}
\newcommand{\ea}{\end{eqnarray}}

\begin{document}

\begin{frontmatter}

\title{The Algebraic Cluster Model: Structure of $^{16}$O}
\author{R. Bijker}
\address{Instituto de Ciencias Nucleares, 
Universidad Nacional Aut\'onoma de M\'exico, \\
Apartado Postal 70-543, 04510 M\'exico, D.F., M\'exico}
\ead{bijker@nucleares.unam.mx}

\author{F. Iachello}
\address{Center for Theoretical Physics, Sloane Laboratory, Yale University, \\ 
New Haven, CT 06520-8120, U.S.A.}
\ead{francesco.iachello@yale.edu}

\begin{abstract}
We discuss an algebraic treatment of four-body clusters which includes both 
continuous and discrete symmetries. In particular, tetrahedral configurations 
with ${\cal T}_d$ symmetry are analyzed with respect to the energy spectrum, 
transition form factors and $B(EL)$ values. It is concluded that the low-lying 
spectrum of $^{16}$O can be described by four $\alpha$ particles at the vertices 
of a regular tetrahedron, not as a rigid structure but rather a more floppy 
structure with relatively large rotation-vibration interactions and Coriolis forces. 
\end{abstract}



\end{frontmatter}

\section{Introduction}

The binding energy per nucleon for light nuclei shows large oscillations with nucleon number 
with maxima for nuclei with $A=4n$ and $Z=N$, especially for the nuclei $^{4}$He, $^{8}$Be, 
$^{12}$C and $^{16}$O for $n=1$, $2$, $3$ and $4$, respectively, which provides a strong 
indication of the importance of $\alpha$ clustering in these nuclei.  
The phenomenon of clustering in light nuclei has a large history dating back to the 1930's 
with early studies on $\alpha$-cluster models by Wheeler \cite{wheeler}, and Hafstad and Teller 
\cite{Teller}, followed by later work by Dennison \cite{Dennison}, Kameny \cite{Kameny}, 
Brink \cite{Brink1,Brink2} and Robson \cite{Robson1,Review}. 
The connection between the nuclear shell model and the cluster model was studied 
in \cite{wilder}, as well as by the the Japanese school \cite{ikeda,fujiwara,horiuchi1,horiuchi2}.  
A recent review on cluster models can be found in Ref.~\cite{Oertzen}.  

In the last few years, there has been considerable renewed interest in the structure of 
$\alpha$-cluster nuclei, especially for the nucleus $^{12}$C \cite{Hoyle}. The measurement 
of new rotational excitations of both the ground state \cite{Fre07,Kirsebom,Marin} and the Hoyle state 
\cite{Itoh,Freer2,Gai,Freer4} has stimulated a large effort to understand the structure of $^{12}$C 
ranging from studies based on antisymmetrized molecular dynamics (AMD) \cite{AMD}, 
fermionic molecular dynamics (FMD) \cite{FMD}, BEC-like cluster model \cite{BEC}, 
{\it ab initio} no-core shell model \cite{NCSM}, lattice EFT \cite{EFT1,EFT2}, no-core symplectic model 
\cite{Draayer} and the algebraic cluster model (ACM) \cite{Marin,ACM1,ACM2,TdO16}. 

The aim of this paper is to develop the ACM for four-body cluster systems and especially to 
discuss an application to the nucleus $^{16}$O as a cluster of four $\alpha$-particles.  
The ACM provides an algebraic treatment of the relative motion of the clusters 
in which eigenvalues and eigenfunctions are obtained by matrix diagonalization 
instead of by solving a set of coupled differential equations.  
Algebraic models have found useful applications both in many- and in few-body 
systems. As an example we mention the interacting boson model (IBM), which has been 
very successful in the description of collective states in nuclei \cite{ibm}. 
Its dynamical symmetries correspond to the quadrupole vibrator \cite{u5}, 
the axially symmetric rotor \cite{su3} and the $\gamma$-unstable rotor \cite{so6} 
in a geometrical description. 
The first extension to few-body systems was the vibron model \cite{vibron}
which was introduced to describe vibrational and rotational excitations in diatomic 
molecules \cite{cpl}. The dynamical symmetries of the vibron model correspond to 
the (an)harmonic oscillator and the Morse oscillator. 

The general procedure is to introduce a $U(k+1)$ spectrum generating
algebra for a bound-state problem with $k$ degrees of freedom \cite{FI}. 
For the five quadrupole degrees of freedom in collective nuclei this led 
to the introduction of the $U(6)$ interacting boson model \cite{ibm}. 
Similarly, the $U(4)$ vibron model was proposed to describe the dynamics of 
the three dipole degrees of freedom of the relative motion of two objects, 
{\it e.g.} two atoms in a diatomic molecule \cite{vibron}, two clusters in 
a nuclear cluster model \cite{cluster1,cluster2,cluster3}, or quark-antiquark configurations 
in mesons \cite{meson1,meson2}. For three-body cluster systems the algebra is $U(7)$. 
The $U(7)$ algebraic cluster model was developed originally to describe the relative motion of the 
three constituent quarks in baryons ($qqq$) \cite{BIL1,BIL2}, but has also found applications 
in molecular physics (H$_3^+$) \cite{BDL,BL} and nuclear physics ($^{12}$C as a cluster 
of three $\alpha$ particles) \cite{ACM1,ACM2}.  

This paper is organized as follows. In Sects.~2-3, we introduce the algebraic cluster model 
for four-body systems in terms of a $U(10)$ spectrum-generating algebra, followed by 
a discussion of the permutation symmetry for the case of four identical objects. 
In Sect.~4, we show that within $U(10)$ it is possible to provide a description of a spherical top 
with tetrahedral symmetry in which all vibrational and rotational degrees of freedom are present 
from the outset. In Sect.~5, we discuss how to calculate transition probabilities in the ACM 
and we apply the $U(10)$ model in the limit of the spherical top to the 
description of the nucleus $^{16}$O as a cluster of four $\alpha$-particles, 
and show evidence for ${\cal T}_d$ symmetry in this nucleus. 

A preliminary account of part of this work has appeared in Refs.~\cite{TdO16,RB1,RB2,RB3}. 
Here we develop the $U(10)$ ACM in more detail, and especially we present a detailed 
analysis of the spectroscopic properties of $^{16}$O including energies, transition 
rates and electromagnetic form factors, and show that $^{16}$O can be, to a good approximation, 
described by a $4\alpha$ configuration with ${\cal T}_d$ symmetry. 

The development of the $U(10)$ ACM described here opens the way for applications 
to systems with generic geometric configurations as mentioned in the final Sect.~7. 

\section{Algebraic Cluster Model}

\begin{figure}
\centering
\setlength{\unitlength}{2pt}
\begin{picture}(120,100)(0,20)
\thicklines
\put( 50, 30) {\circle*{5}} 
\put( 30, 50) {\circle*{5}}
\put( 80, 50) {\circle*{5}}
\put( 60, 90) {\circle*{5}}
\put( 50, 30) {\vector(-1, 1){18}} 
\thinlines
\multiput( 30, 50)( 3, 0){18}{\circle*{1}}
\put( 80, 50) {\vector(-4,-1){40}}
\put( 60, 90) {\vector( 0,-1){45}}
\thicklines
\put( 50, 30) {\line(-1, 1){20}}
\put( 50, 30) {\line( 3, 2){30}}
\put( 60, 90) {\line(-3,-4){30}}
\put( 60, 90) {\line( 1,-2){20}}
\put( 60, 90) {\line(-1,-6){10}}
\put( 35, 35) {$\vec{\rho}$}
\put( 55, 37) {$\vec{\lambda}$}
\put( 63, 60) {$\vec{\eta}$}
\put( 20, 50) {$1$}
\put( 50, 20) {$2$}
\put( 85, 50) {$3$}
\put( 60, 95) {$4$}
\end{picture}
\caption[]
{Jacobi coordinates in a tetrahedral configuration}
\label{shape}
\end{figure}
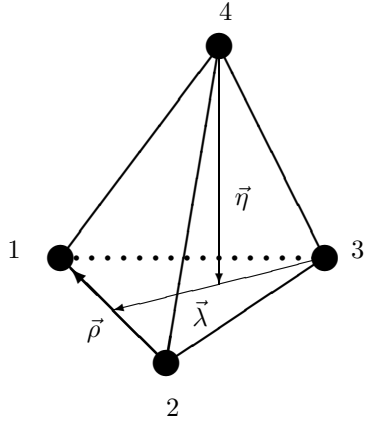

The Algebraic Cluster Model (ACM) is a model designed to describe the relative motion 
of a cluster system. We start by introducing the relative Jacobi coordinates for a four-body 
system, see Fig.~\ref{shape}, 
\ba
\vec{\rho} &=& \left( \vec{r}_{1} - \vec{r}_{2} \right) /\sqrt{2} ~,  
\nonumber\\
\vec{\lambda} &=& \left( \vec{r}_{1} + \vec{r}_{2} - 2\vec{r}_{3} \right)
/\sqrt{6} ~, 
\nonumber\\
\vec{\eta} &=& \left( \vec{r}_{1} + \vec{r}_{2} + \vec{r}_{3} - 3\vec{r}_{4} \right)
/\sqrt{12} ~,  
\label{Jacobi}
\ea
together with their conjugate momenta. Here $\vec{r}_i$ ($i=1,\ldots,4$) represent the 
coordinates of the four constituent particles. Instead of a formulation in terms of coordinates and 
momenta we use the method of bosonic quantization which consists in quantizing the Jacobi 
coordinates and momenta with vector boson operators and adding an additional 
scalar boson 
\ba 
b^{\dagger}_{\rho m} ~, \; b^{\dagger}_{\lambda m} ~, \; b^{\dagger}_{\eta m} ~, \; 
s^{\dagger} ~, \hspace{1cm} (m=-1,0,1) ~,
\label{bb}
\ea
altoghether denoted as $c^{\dagger}_i$ with $i=1.\ldots,10$, 
under the constraint that the Hamiltonian commutes with the number operator
\ba
\hat{N} \;=\; \sum_i c^{\dagger}_i c_i \;=\; s^{\dagger}s + \sum_{m} \left( b_{\rho m}^{\dagger} b_{\rho m} 
+ b_{\lambda m}^{\dagger} b_{\lambda m} + b_{\eta m}^{\dagger} b_{\eta m} \right) ~, 
\label{number}
\ea
{\it i.e.} the total number of bosons $N=n_s+n_{\rho}+n_{\lambda}+n_{\eta}$ is conserved.  
The set of 100 bilinear products of creation and annihilation operators $G_{ij}=c^{\dagger}_i c_j$ 
spans the Lie algebra of $U(10)$. All operators of interest, such as the Hamiltonian and 
electromagnetic transition operators, are expressed in terms of elements of this algebra.  
As an example, the one- and two-body Hamiltonian is given by 
\ba
H \;=\; \sum_{ij} \epsilon_{ij} G_{ij} + \sum_{ijkl} v_{ijkl} G_{ij} G_{kl} ~. 
\ea
The model space of the ACM is spanned by the symmetric irreducible representation $[N]$ of $U(10)$ 
which contains the oscillator shells with $n=n_{\rho}+n_{\lambda}+n_{\eta}=0,1,\ldots, N$. 
The introduction of the scalar boson makes it possible to investigate the dynamics 
of three vector degrees of freedom including situations in which there is a mixing of 
oscillator shells. It can be seen as a convenient way to compactify the infinite 
dimensional space of the harmonic oscillator to a finite-dimensional space.  

\section{Permutation symmetry}

In case of identical clusters, the Hamiltonian should be invariant under their permutation 
and, as a consequence, the states would transform according to the representations of the 
corresponding permutation group. 

For four identical objects, as for example for X$_4$ molecules or $4\alpha$ clusters,  
the Hamiltonian has to be invariant under the permuation group $S_4$. The permutation 
symmetry of four identical objects is determined by the transposition $P(12)$ and the 
cyclic permutation $P(1234)$ \cite{KM}. All other permutations can be expressed in 
terms of these two elementary ones. The transformation properties under $S_4$ of all
operators in the model follow from those of the building blocks. 
Algebraically, the transposition and cyclic permutation can be expressed in terms of 
the generators $b^{\dagger}_{i} b_{j} \equiv \sum_{m} b^{\dagger}_{i,m} b_{j,m}$ that act in 
index space ($i,j=\rho$, $\lambda$, $\eta$). The transposition is given by 
\ba
P(12) \left( \begin{array}{c} s^{\dagger} \\ b^{\dagger}_{\rho} \\ b^{\dagger}_{\lambda} \\ 
b^{\dagger}_{\eta} \end{array} \right) \;=\; U_{\rm tr} 
\left( \begin{array}{c} s^{\dagger} \\ b^{\dagger}_{\rho} \\ b^{\dagger}_{\lambda} \\ 
b^{\dagger}_{\eta} \end{array} \right) U^{-1}_{\rm tr} 
\;=\; \left( \begin{array}{rrrr} 1 & 0 & 0 & 0 \\ 0 & -1 & 0 & 0 \\ 0 & 0 & 1 & 0 \\ 
0 & 0 & 0 & 1 \end{array} \right) \left( \begin{array}{c} s^{\dagger} \\ b^{\dagger}_{\rho} \\ 
b^{\dagger}_{\lambda} \\ b^{\dagger}_{\eta} \end{array} \right) ~, 
\ea
with 
\ba
U_{\rm tr} \;=\; \mbox{e}^{i \pi b^{\dagger}_{\rho} b_{\rho}} ~,  
\label{p12}
\ea 
and the cyclic permutation by
\ba
P(1234) \left( \begin{array}{c} s^{\dagger} \\ b^{\dagger}_{\rho} \\ b^{\dagger}_{\lambda} \\ 
b^{\dagger}_{\eta} \end{array} \right) &=& U_{\rm cycl} 
\left( \begin{array}{c} s^{\dagger} \\ b^{\dagger}_{\rho} \\ b^{\dagger}_{\lambda} \\ 
b^{\dagger}_{\eta} \end{array} \right) U^{-1}_{\rm cycl}
\nonumber\\ 
&=& \left( \begin{array}{cccc} 1 & 0 & 0 & 0 \\ 0 & -\frac{1}{2} & \frac{\sqrt{3}}{2} & 0 \\ 
0 & -\frac{1}{2\sqrt{3}} & -\frac{1}{6} & \frac{\sqrt{8}}{3} \\ 
0 & -\frac{\sqrt{2}}{\sqrt{3}} & -\frac{\sqrt{2}}{3} & -\frac{1}{3} \end{array} \right)  
\left( \begin{array}{c} s^{\dagger} \\ b^{\dagger}_{\rho} \\ b^{\dagger}_{\lambda} \\ 
b^{\dagger}_{\eta} \end{array} \right) ~, 
\ea
with
\ba
U_{\rm cycl} \;=\; \mbox{e}^{i \pi (b^{\dagger}_{\rho} b_{\rho} 
+ b^{\dagger}_{\lambda} b_{\lambda}+b^{\dagger}_{\eta} b_{\eta})} \, 
\mbox{e}^{\theta_1 (b^{\dagger}_{\rho} b_{\lambda} - b^{\dagger}_{\lambda} b_{\rho})} \, 
\mbox{e}^{\theta_2 (b^{\dagger}_{\lambda} b_{\eta} - b^{\dagger}_{\eta} b_{\lambda})} ~, 
\label{p1234}
\ea
and $\theta_1=\arctan \sqrt{3}$ and $\theta_2=\arctan \sqrt{8}$. 
The scalar boson, $s^{\dagger}$, transforms as the symmetric representation $[4]$, whereas the 
three vector Jacobi bosons, $b^{\dagger}_{\rho}$, $b^{\dagger}_{\lambda}$ and $b^{\dagger}_{\eta}$, 
transform as the three components of the mixed symmetry representation $[31]$. 

There are five different symmetry classes for the permutation of four objects. 
Since $S_4$ is isomorphic to the tetrahedral group ${\cal T}_d$, the irreducible 
representations can also be labeled by those of the tetrahedral group 
\ba
\, [4] & \sim & A_1 ~,
\nonumber\\
\, [31] & \sim & F_2 ~, 
\nonumber\\   
\, [22] & \sim & E ~,
\nonumber\\ 
\, [211] & \sim & F_1 ~,
\nonumber\\
\, [1111] & \sim & A_2 ~.
\ea
In the remainder of this article we will use the irreducible representations of ${\cal T}_d$ 
to indicate the tensorial character under the permutation group. 

Next, one can use the multiplication rules for ${\cal T}_d$ to construct physical operators with the 
appropriate symmetry properties. For example, for the bilinear products of the three vector 
Jacobi bosons, one finds  
\ba
F_2 \otimes F_2 \;=\; A_1 \oplus F_2 \oplus F_1 \oplus E ~.
\ea
It is convenient to express the 100 elements of the algebra of $U(10)$ in terms of tensor 
operators under both the rotation group $SO(3)$ and the tetrahedral group ${\cal T}_d \sim S_4$ as 
\ba
\begin{array}{lcl}
A_1 && \hspace{0.5cm} \hat{n}_s \;=\; (s^{\dagger} \tilde{s})^{(0)}_0 \\ & & \\
F_2 && \left\{ \begin{array}{l}
\hat D_{\rho} \;=\; (b^{\dagger}_{\rho} \tilde{s} -
s^{\dagger} \tilde{b}_{\rho})^{(1)} \\ 
\hat D_{\lambda} \;=\; (b^{\dagger}_{\lambda} \tilde{s} -
s^{\dagger} \tilde{b}_{\lambda})^{(1)} \\ 
\hat D_{\eta} \;=\; (b^{\dagger}_{\eta} \tilde{s} -
s^{\dagger} \tilde{b}_{\eta})^{(1)} \end{array} \right.\\ & & \\
F_2 && \left\{ \begin{array}{l}
\hat A_{\rho} \;=\; i \, (b^{\dagger}_{\rho} \tilde{s} +
s^{\dagger} \tilde{b}_{\rho})^{(1)} \\
\hat A_{\lambda} \;=\; i \, (b^{\dagger}_{\lambda} \tilde{s} +
s^{\dagger} \tilde{b}_{\lambda})^{(1)} \\
\hat A_{\eta} \;=\; i \, (b^{\dagger}_{\eta} \tilde{s} +
s^{\dagger} \tilde{b}_{\eta})^{(1)} \end{array} \right.\\ & & \\
A_1 && \hspace{0.5cm} \hat B^{(l)} \;=\; ( b^{\dagger}_{\rho} \tilde{b}_{\rho}
+ b^{\dagger}_{\lambda} \tilde{b}_{\lambda} 
+ b^{\dagger}_{\eta} \tilde{b}_{\eta})^{(l)} \\ & & \\
E && \left\{ \begin{array}{l}
\hat C^{(l)}_{\rho} \;=\; ( b^{\dagger}_{\rho} \tilde{b}_{\lambda} 
+ b^{\dagger}_{\lambda} \tilde{b}_{\rho} )^{(l)} 
- \sqrt{2}( b^{\dagger}_{\rho} \tilde{b}_{\eta} 
+ b^{\dagger}_{\eta} \tilde{b}_{\rho} )^{(l)} \\ 
\hat C^{(l)}_{\lambda} \;=\; ( b^{\dagger}_{\rho} \tilde{b}_{\rho}
- b^{\dagger}_{\lambda} \tilde{b}_{\lambda} )^{(l)} 
- \sqrt{2} ( b^{\dagger}_{\lambda} \tilde{b}_{\eta}
+ b^{\dagger}_{\eta} \tilde{b}_{\lambda} )^{(l)} \end{array} \right.\\ & & \\
F_2 && \left\{ \begin{array}{l}
\hat G^{(l)}_{\rho} \;=\; \sqrt{2}( b^{\dagger}_{\rho} \tilde{b}_{\lambda} 
+ b^{\dagger}_{\lambda} \tilde{b}_{\rho} )^{(l)} 
+ ( b^{\dagger}_{\rho} \tilde{b}_{\eta} 
+ b^{\dagger}_{\eta} \tilde{b}_{\rho} )^{(l)} \\
\hat G^{(l)}_{\lambda} \;=\; \sqrt{2} ( b^{\dagger}_{\rho} \tilde{b}_{\rho}
- b^{\dagger}_{\lambda} \tilde{b}_{\lambda} )^{(l)} 
+ ( b^{\dagger}_{\lambda} \tilde{b}_{\eta}
+ b^{\dagger}_{\eta} \tilde{b}_{\lambda} )^{(l)} \\
\hat G^{(l)}_{\eta} \;=\; ( b^{\dagger}_{\rho} \tilde{b}_{\rho}
+ b^{\dagger}_{\lambda} \tilde{b}_{\lambda}  
- 2 b^{\dagger}_{\eta} \tilde{b}_{\eta} )^{(l)} \end{array} \right.\\ & & \\
F_1 && \left\{ \begin{array}{l}
\hat K^{(l)}_{\rho} \;=\; -i \, 
( b^{\dagger}_{\lambda} \tilde{b}_{\eta} 
- b^{\dagger}_{\eta} \tilde{b}_{\lambda} )^{(l)} \\
\hat K^{(l)}_{\lambda} \;=\; -i \, 
( b^{\dagger}_{\eta} \tilde{b}_{\rho} 
- b^{\dagger}_{\rho} \tilde{b}_{\eta} )^{(l)} \\
\hat K^{(l)}_{\eta} \;=\; -i \, 
( b^{\dagger}_{\rho} \tilde{b}_{\lambda} 
- b^{\dagger}_{\lambda} \tilde{b}_{\rho} )^{(l)} \end{array} \right.\\
\label{gen}
\end{array}
\ea
with $l=0,1,2$. Here $\tilde{s}=s$ and $\tilde{b}_{k m}=(-1)^{1-m} b_{k,-m}$ where  
$k$ denotes the three Jacobi coordinates $\rho$, $\lambda$, $\eta$. 

The invariance under tetrahedral symmetry imposes strong constraints on the general 
Hamiltonian. The most general one- and two-body Hamiltonian that is rotationally 
invariant, conserves parity as well as the total number of bosons, and in addition 
is scalar under the tetrahedral group ${\cal T}_d \sim S_4$, is given by 
\ba
H &=& \epsilon_{0} \, s^{\dagger} \tilde{s}
- \epsilon_{1} \, (b_{\rho}^{\dagger} \cdot \tilde{b}_{\rho} 
+ b_{\lambda}^{\dagger} \cdot \tilde{b}_{\lambda}  
+ b_{\eta}^{\dagger} \cdot \tilde{b}_{\eta})
\nonumber\\ 
&& + u_0 \, s^{\dagger} s^{\dagger} \tilde{s} \tilde{s}  
- u_1 \, s^{\dagger} ( b_{\rho}^{\dagger} \cdot \tilde{b}_{\rho} 
+ b_{\lambda}^{\dagger} \cdot \tilde{b}_{\lambda}  
+ b_{\eta}^{\dagger} \cdot \tilde{b}_{\eta} ) \tilde{s} 
\nonumber\\
&& + v_0 \, \left[ ( b_{\rho}^{\dagger} \cdot b_{\rho}^{\dagger} 
+ b_{\lambda}^{\dagger} \cdot b_{\lambda}^{\dagger}  
+ b_{\eta}^{\dagger} \cdot b_{\eta}^{\dagger} ) \tilde{s} \tilde{s} + {\rm h.c.} \right]
\nonumber\\
&& + \sum_{L=0,2} a_{L} \, \left[ [ 2 b_{\rho}^{\dagger} b_{\eta}^{\dagger}
+ 2\sqrt{2} \, b_{\rho}^{\dagger} b_{\lambda}^{\dagger} ]^{(L)} 
\cdot [ {\rm h.c.} ]^{(L)} \right.
\nonumber\\
&& \hspace{2cm} + [ 2 b_{\lambda}^{\dagger} b_{\eta}^{\dagger} 
+ \sqrt{2} \, ( b_{\rho}^{\dagger} b^{\dagger}_{\rho}  
- b_{\lambda}^{\dagger} b_{\lambda}^{\dagger} ) ]^{(L)} \cdot [ {\rm h.c.} ]^{(L)} 
\nonumber\\
&& \hspace{2cm} \left. + [ b_{\rho}^{\dagger} b_{\rho}^{\dagger} 
     + b_{\lambda}^{\dagger} b_{\lambda}^{\dagger} 
   - 2 b_{\eta}^{\dagger} b_{\eta}^{\dagger} ]^{(L)} \cdot [ {\rm h.c.} ]^{(L)} \right]
\nonumber\\
&& + \sum_{L=0,2} c_{L} \, \left[
[ -2\sqrt{2} \, b_{\rho}^{\dagger} b_{\eta}^{\dagger}
+ 2 b_{\rho}^{\dagger} b_{\lambda}^{\dagger} ]^{(L)} \cdot [ {\rm h.c.} ]^{(L)} \right.
\nonumber\\
&& \hspace{2cm} + \left. [ -2\sqrt{2} \, b_{\lambda}^{\dagger} b_{\eta}^{\dagger} 
+ ( b_{\rho}^{\dagger} b_{\rho}^{\dagger} 
- b_{\lambda}^{\dagger} b_{\lambda}^{\dagger} ) ]^{(L)} \cdot [ {\rm h.c.} ]^{(L)} \right]
\nonumber\\
&& + c_1 \, \left[ 
  ( b_{\rho}^{\dagger} b_{\lambda}^{\dagger} )^{(1)} \cdot 
  ( \tilde{b}_{\lambda} \tilde{b}_{\rho} )^{(1)} 
+ ( b_{\lambda}^{\dagger} b_{\eta}^{\dagger} )^{(1)} \cdot 
  ( \tilde{b}_{\eta} \tilde{b}_{\lambda} )^{(1)} 
+ ( b_{\eta}^{\dagger} b_{\rho}^{\dagger} )^{(1)} \cdot 
  ( \tilde{b}_{\rho} \tilde{b}_{\eta} )^{(1)} \right]
\nonumber\\
&& + \sum_{L=0,2} d_{L} \,
( b_{\rho}^{\dagger} b_{\rho}^{\dagger} 
+ b_{\lambda}^{\dagger} b_{\lambda}^{\dagger} 
+ b_{\eta}^{\dagger} b_{\eta}^{\dagger} )^{(L)} \cdot ( {\rm h.c.} )^{(L)} ~.
\label{HS4}
\ea
By construction, the wave 
functions are characterized by the total number of bosons $N$, angular momentum $L$ 
and parity $P$, and their transformation property $t$ under the tetrahedral group. 
Since we do not consider internal excitations of the clusters, the four-body wave 
functions arise solely from the relative motion and are symmetric with $t=A_1$. 

The relation between harmonic oscillators and permutation symmetry was studied 
by Kramer and Moshinsky \cite{KM}. In the present case, we wish to study the 
properties of the $U(10)$ algebraic cluster model for any (large) number of oscillator 
quanta with the possibility of mixing between different oscillator shells. Therefore, 
we prefer to generate a set of basis states with good permutation symmetry numerically 
by diagonalization of ${\cal T}_d \sim S_4$ invariant interactions. The permutation 
symmetry $t$ of a given eigenfunction can then be determined from the transformation 
properties under the transposition $P(12)$ and the cyclic permutation $P(1234)$ 
\cite{KM}. In practice, the wave functions are obtained numerically by diagonalization, 
and hence are determined up to a sign. The relative phases of the degenerate representations, 
the two-dimensional $E$, and the three-dimensional $F_2$ and $F_1$, can be determined from 
the off-diagonal matrix elements of $P(1234)$ \cite{RB3}. 

\section{Special solutions}

In general, the eigenvalues and corresponding eigenvectors are obtained numerically by 
diagonalizing the Hamiltonian of Eq.~(\ref{HS4}) in a coupled harmonic oscillator basis. 
However, there are special limiting cases of the Hamiltonian of Eq.~(\ref{HS4}), in which 
the energy spectra can be obtained in closed form. These special cases are called 
dynamical symmetries and arise whenever the Hamiltonian is expressed in terms of the Casimir 
invariants of a chain of subalgebras of $U(10)$. Two examples of dynamical symmetries of the 
$S_4 \sim {\cal T}_d$ invariant Hamiltonian correspond to the group lattice 
\ba
U(10) \supset \left\{ \begin{array}{c} U(9) \\ \\ SO(10) \end{array} \right\} \supset SO(9) 
\supset {\cal SO}(3) \otimes SO(3) ~,
\label{lattice}
\ea 
where ${\cal SO}(3)$ denotes the angular momentum group in coordinate space and $SO(3)$ the 
angular momentum in index space. 
These dynamical symmetries were shown to correspond to the nine-dimensional (an)harmonic 
oscillator and the nine-dimensional deformed oscillator, respectively \cite{RB1,RB2,RB3}.

In the following, we discuss in more detail the case of the spherical top. Although it does 
not correspond to a dynamical symmetry, approximate solutions can still be obtained in the 
large $N$ limit, which subsequently will be used to analyze and interpret the cluster states 
in $^{16}$O. An interesting limiting case of the general ${\cal T}_d$ invariant Hamiltonian 
of Eq.~(\ref{HS4}) is provided by \cite{TdO16}
\ba
H_{3,\rm vib} &=& \xi_{1} \, (R^{2} \,s^{\dagger} s^{\dagger} 
- b_{\rho}^{\dagger} \cdot b_{\rho}^{\dagger} 
- b_{\lambda }^{\dagger} \cdot b_{\lambda}^{\dagger} 
- b_{\eta}^{\dagger} \cdot b_{\eta}^{\dagger}) \, ( {\rm h.c.} )  
\nonumber\\
&& + \xi_2 \, \left[ ( -2\sqrt{2} \, b_{\rho}^{\dagger} \cdot b_{\eta}^{\dagger} 
+ 2 b_{\rho}^{\dagger} \cdot b_{\lambda}^{\dagger} ) \, ( {\rm h.c.} ) \right.
\nonumber\\
&& \hspace{2cm} + \left. ( -2\sqrt{2} \, b_{\lambda}^{\dagger} \cdot b_{\eta}^{\dagger} 
+ ( b_{\rho}^{\dagger} \cdot b_{\rho}^{\dagger} 
- b_{\lambda}^{\dagger} \cdot b_{\lambda}^{\dagger} )) \, ( {\rm h.c.} ) \right]
\nonumber\\
&& + \xi_3 \, \left[ ( 2 b_{\rho}^{\dagger} \cdot b_{\eta}^{\dagger}
+ 2\sqrt{2} \, b_{\rho}^{\dagger} \cdot b_{\lambda}^{\dagger} ) \, ( {\rm h.c.} ) \right.
\nonumber\\
&& \hspace{2cm} + ( 2 b_{\lambda}^{\dagger} \cdot b_{\eta}^{\dagger} 
+ \sqrt{2} \, ( b_{\rho}^{\dagger} \cdot b^{\dagger}_{\rho}  
- b_{\lambda}^{\dagger} \cdot b_{\lambda}^{\dagger} )) \, ( {\rm h.c.} )
\nonumber\\
&& \hspace{2cm} \left. + ( b_{\rho}^{\dagger} \cdot b_{\rho}^{\dagger} 
     + b_{\lambda}^{\dagger} \cdot b_{\lambda}^{\dagger} 
   - 2 b_{\eta}^{\dagger} \cdot b_{\eta}^{\dagger} ) \, ( {\rm h.c.} ) \right] ~. 
\label{h3vib}
\ea
This Hamiltonian is a particular combination of the $U(9)$ and $SO(10)$ dynamical symmetries. 
For $R^{2}=0$, this Hamiltonian has $U(10) \supset U(9)$ symmetry and corresponds to a 
nine-dimensional anharmonic oscillator, whereas for $R^{2}=1$ and $\xi_2=\xi_3=0$ it has 
$U(10) \supset SO(10)$ symmetry and corresponds to a deformed oscillator. For the general case 
with $R^{2} \neq 0$ and $\xi_{1}$, $\xi_{2}$, $\xi_{3}>0$ the energy spectrum cannot be obtained 
in closed analytic form. Even though the Hamiltonian of Eq.~(\ref{h3vib}) does not correspond to 
a dynamical symmetry, an approximate energy formula can still be derived by studying the classical 
limit of Eq.~(\ref{h3vib}). 

The classical limit of the ACM Hamiltonian is defined by the coherent state expectation value 
\ba
H_{\rm cl} \;=\; \frac{1}{N} 
\langle N;\vec{\alpha}_{\rho},\vec{\alpha}_{\lambda},\vec{\alpha}_{\eta} \mid \,: H :\, 
\mid N;\vec{\alpha}_{\rho},\vec{\alpha}_{\lambda},\vec{\alpha}_{\eta} \rangle ~,
\ea
where the coherent state has the form of a condensate wave function
\ba
|N;\vec{\alpha}_{\rho},\vec{\alpha}_{\lambda},\vec{\alpha}_{\eta} \rangle \;=\; 
\frac{1}{\sqrt{N!}} (b_{c}^{\dagger})^{N}\,|0\rangle ~. 
\label{coherent}
\ea
The condensate boson $b_c^{\dagger}$ is parametrized in terms of nine complex variables 
corresponding to the three vector coordinates and their conjugate momenta  
\ba
b_{c}^{\dagger} \;=\; \sqrt{1-\vec{\alpha}_{\rho} \cdot \vec{\alpha}_{\rho}^{\,\ast}
-\vec{\alpha}_{\lambda} \cdot \vec{\alpha}_{\lambda}^{\,\ast} 
-\vec{\alpha}_{\eta} \cdot \vec{\alpha}_{\eta}^{\,\ast}} 
\; s^{\dagger} + \vec{\alpha}_{\rho} \cdot \vec{b}_{\rho}^{\,\dagger } 
+ \vec{\alpha}_{\lambda} \cdot \vec{b}_{\lambda}^{\,\dagger} 
+ \vec{\alpha}_{\eta} \cdot \vec{b}_{\eta}^{\,\dagger} ~.
\label{bc}
\ea
For the geometrical analysis of the ACM Hamiltonian it is convenient to make a 
transformation to spherical coordinates and momenta \cite{onno,Levit} 
\ba
\alpha _{k,\mu } \;=\; \frac{1}{\sqrt{2}}\sum_{\nu }{\cal D}_{\mu \nu }^{(1)}(\phi
_{k},\theta _{k},0)\,\beta _{k,\nu }~,
\ea
with 
\ba
\left( \begin{array}{c} \beta _{k,1} \\ \beta _{k,0} \\ \beta _{k,-1} \end{array} \right) 
\;=\; \left( \begin{array}{c}
\,[-p_{\phi _{k}}/\sin \theta _{k}-ip_{\theta _{k}}]/q_{k}\sqrt{2} \\ q_{k}+ip_{k} \\ 
\,[-p_{\phi _{k}}/\sin \theta _{k}+ip_{\theta _{k}}]/q_{k}\sqrt{2}
\end{array} \right) ~,
\ea
where $k=\rho$, $\lambda$, $\eta$, followed by a change of variables to the hyperspherical 
coordinates $q$, $\zeta$ and $\chi$ 
\ba
q_{\rho} &=& q \sin \zeta \sin \chi ~, 
\nonumber\\
q_{\lambda} &=& q \sin \zeta \cos \chi ~,
\nonumber\\
q_{\eta} &=& q \cos \zeta ~,
\label{hyper}
\ea
and to center-of-mass $\Omega$ and relative angles $\theta_{ij}$ and their conjugate momenta. 
Here $2\theta_{\rho\lambda}$ denotes the relative angle between $\vec{\alpha}_{\rho}$ and 
$\vec{\alpha}_{\lambda}$, and similarly for $2\theta_{\lambda\eta}$ and $2\theta_{\eta\rho}$. 

\begin{figure}
\centering
\vspace{15pt}
\setlength{\unitlength}{1pt}
\begin{picture}(330,240)(0,0)
\thicklines
\put( 50,130) {\circle*{5}} 
\put( 30,150) {\circle*{5}}
\put( 80,150) {\circle*{5}}
\put( 60,190) {\circle*{5}}
\put( 50,130) {\vector( 0,-1){10}} 
\put( 30,150) {\vector(-1, 0){10}}
\put( 80,150) {\vector( 1, 0){10}}
\put( 60,190) {\vector( 0, 1){10}}
\put( 50,130) {\line(-1, 1){20}}
\put( 50,130) {\line( 3, 2){30}}
\put( 60,190) {\line(-3,-4){30}}
\put( 60,190) {\line( 1,-2){20}}
\put( 60,190) {\line(-1,-6){10}}
\put( 10,190) {$v_1(A_1)$}
\multiput( 30,150)(5,0){10}{\circle*{1}}
\put(150,130) {\circle*{5}} 
\put(130,150) {\circle*{5}}
\put(180,150) {\circle*{5}}
\put(160,190) {\circle*{5}}
\put(150,130) {\vector( 2,-1){10}} 
\put(130,150) {\vector(-1, 2){ 5}}
\put(180,150) {\vector( 1,-2){ 5}}
\put(160,190) {\vector(-2, 1){10}}
\put(150,130) {\line(-1, 1){20}}
\put(150,130) {\line( 3, 2){30}}
\put(160,190) {\line(-3,-4){30}}
\put(160,190) {\line( 1,-2){20}}
\put(160,190) {\line(-1,-6){10}}
\put(110,190) {$v_{2a}(E)$}
\multiput(130,150)(5,0){10}{\circle*{1}}
\put(250,130) {\circle*{5}} 
\put(230,150) {\circle*{5}}
\put(280,150) {\circle*{5}}
\put(260,190) {\circle*{5}}
\put(250,130) {\vector( 1, 2){ 7}} 
\put(230,150) {\vector( 3, 2){10}}
\put(280,150) {\vector(-3,-4){10}}
\put(260,190) {\vector(-3, 1){10}}
\put(250,130) {\line(-1, 1){20}}
\put(250,130) {\line( 3, 2){30}}
\put(260,190) {\line(-3,-4){30}}
\put(260,190) {\line( 1,-2){20}}
\put(260,190) {\line(-1,-6){10}}
\put(210,190) {$v_{2b}(E)$}
\multiput(230,150)(5,0){10}{\circle*{1}}
\put( 50, 30) {\circle*{5}} 
\put( 30, 50) {\circle*{5}}
\put( 80, 50) {\circle*{5}}
\put( 60, 90) {\circle*{5}}
\put( 50, 30) {\vector(-1, 1){10}} 
\put( 30, 50) {\vector( 1,-1){10}}
\put( 80, 50) {\vector( 1,-2){ 5}}
\put( 60, 90) {\vector(-1, 2){ 5}}
\put( 50, 30) {\line(-1, 1){20}}
\put( 50, 30) {\line( 3, 2){30}}
\put( 60, 90) {\line(-3,-4){30}}
\put( 60, 90) {\line( 1,-2){20}}
\put( 60, 90) {\line(-1,-6){10}}
\put( 10, 90) {$v_{3a}(F_2)$}
\multiput( 30, 50)(5,0){10}{\circle*{1}}
\put(150, 30) {\circle*{5}} 
\put(130, 50) {\circle*{5}}
\put(180, 50) {\circle*{5}}
\put(160, 90) {\circle*{5}}
\put(150, 30) {\vector( 0,-1){10}} 
\put(130, 50) {\vector( 1, 0){10}}
\put(180, 50) {\vector(-1, 0){10}}
\put(160, 90) {\vector( 0, 1){10}}
\put(150, 30) {\line(-1, 1){20}}
\put(150, 30) {\line( 3, 2){30}}
\put(160, 90) {\line(-3,-4){30}}
\put(160, 90) {\line( 1,-2){20}}
\put(160, 90) {\line(-1,-6){10}}
\put(110, 90) {$v_{3b}(F_2)$}
\multiput(130, 50)(5,0){10}{\circle*{1}}
\put(250, 30) {\circle*{5}} 
\put(230, 50) {\circle*{5}}
\put(280, 50) {\circle*{5}}
\put(260, 90) {\circle*{5}}
\put(250, 30) {\vector( 3, 2){10}} 
\put(230, 50) {\vector(-3,-4){ 7}}
\put(280, 50) {\vector(-3,-2){10}}
\put(260, 90) {\vector( 3, 4){ 7}}
\put(250, 30) {\line(-1, 1){20}}
\put(250, 30) {\line( 3, 2){30}}
\put(260, 90) {\line(-3,-4){30}}
\put(260, 90) {\line( 1,-2){20}}
\put(260, 90) {\line(-1,-6){10}}
\put(210, 90) {$v_{3c}(F_2)$}
\multiput(230, 50)(5,0){10}{\circle*{1}}
\end{picture}
\caption[]{Fundamental vibrations of a  tetrahedral configuration (point group ${\cal T}_d$).}
\label{fundvib}
\end{figure}
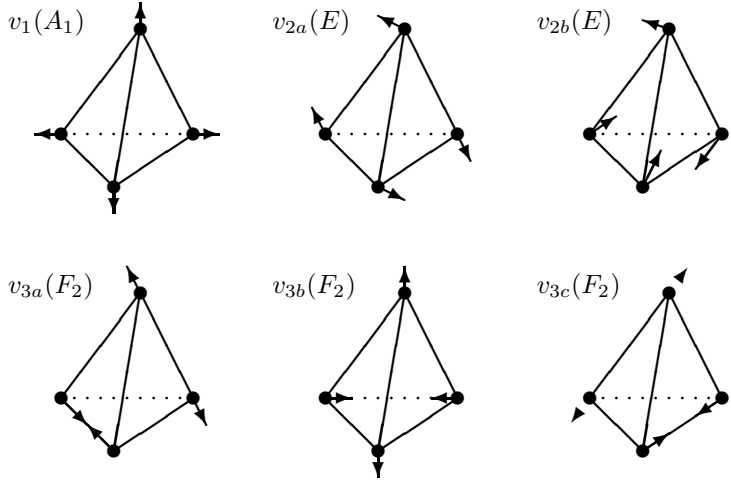

The potential energy surface associated with $H_{3,\rm vib}$ is obtained by setting all momenta 
equal to zero in the general expression for the classical limit. The equilibrium configuration 
corresponds to coordinates that have equal length ($q_{\rho,0}=q_{\lambda,0}=q_{\eta,0}=q_0/\sqrt{3}$)  
\ba
q_{0} \;=\; \sqrt{2R^{2}/(1+R^{2})} ~, \hspace{1cm} \zeta_0 \;=\; \arctan \sqrt{2} ~,  
\hspace{1cm} \chi_0 \;=\; \pi/4 ~,
\label{radii}
\ea
and are mutually perpendicular
\ba  
\theta_{\rho\lambda,0} \;=\; \theta_{\lambda\eta,0} \;=\; \theta_{\eta\rho,0} \;=\; \pi/4 ~.
\label{angles}
\ea
In the limit of small oscillations around the equilibrium shape, the intrinsic degrees of freedom 
decouple and become harmonic. To leading order in $N$ one finds the vibrational energy spectrum of 
a spherical top with tetrahedral symmetry \cite{Herzberg} 
\ba
E_{3,{\rm vib}} \;=\; \omega_{1}(v_{1}+\frac{1}{2}) 
+ \omega_{2}(v_{2}+1) + \omega_{3}(v_{3}+\frac{3}{2}) ~.  
\label{e3vib}
\ea
The frequencies are related to the $\xi$ coefficients in the vibrational Hamiltonian of Eq.~(\ref{h3vib}) 
\ba
\omega_{1} \;=\; 4NR^{2} \xi_{1} ~,  \hspace{1cm} 
\omega_{2} \;=\; \frac{8NR^{2}}{1+R^{2}} \xi_{2} ~, \hspace{1cm} 
\omega_{3} \;=\; \frac{8NR^{2}}{1+R^{2}} \xi_{3} ~.
\ea
Here $v_{1}$ represents the vibrational quantum number for a symmetric stretching $A_1$ vibration, 
$v_2=v_{2a}+v_{2b}$ denotes a doubly degenerate $E$ vibration, and $v_3=v_{3a}+v_{3b}+v_{3c}$ a 
three-fold degenerate $F_2$ vibration (see Fig.~\ref{fundvib}). For rigid configurations, $R^2=1$, 
and $\omega_i=4N \xi_i$ with $i=1,2,3$.  

Next we consider the rotational part of the Hamiltonian 
\ba
H_{3,\rm rot} &=& \kappa_1 \, \vec{L} \cdot \vec{L} 
+ \kappa_2 \, (\vec{L} \cdot \vec{L} - \vec{I} \cdot \vec{I})^2 ~,
\label{h3rot}
\ea
where $\vec{L}$ and $\vec{I}$ denote the angular momentum in coordinate space and 
index space, respectively,  
\ba
L_m &=& \sqrt{2} \, \hat{B}^{(1)}_m ~, \hspace{1cm} m=-1,0,1
\nonumber\\ 
I_j &=& \sqrt{3} \, \hat{K}^{(0)}_j ~, \hspace{1cm} j=\rho, \lambda, \eta
\ea
Whereas the angular momentum $L$ is an exact symmetry of the ${\cal T}_d \sim S_{4}$ 
invariant Hamiltonian of Eq.~(\ref{HS4}), the angular momentum in index space $I$ in 
general does not commute with the Hamiltonian. Only if $a_L=c_L$ in Eq.~(\ref{HS4}) and 
$\xi_2=\xi_3$ in Eq.~(\ref{h3vib}), does $I$ become a good quantum number. The rotational 
excitations of the ground state vibrational band of $H_3$ with $(v_1 v_2 v_3)=(000)$ are 
characterized by $L=I$. This property comes from the fact that the 
operator $\vec{L} \cdot \vec{L} - \vec{I} \cdot \vec{I}$ annihilates the coherent 
(or intrinisic) state corresponding to the rigid equilibrium shape of 
Eqs.~(\ref{radii},\ref{angles}). As a consequence, the rotational energies of the 
ground state band are given by $\kappa_{1} L(L+1)$. 
\ba
E_{3,\rm rot} &=& \kappa_1 \, L(L+1) ~.   
\label{e3rot}
\ea

\begin{figure}
\centering
\setlength{\unitlength}{1.0pt} 
\begin{picture}(320,280)(0,0)
\thinlines
\put (  0,  0) {\line(1,0){310}}
\put (  0,280) {\line(1,0){310}}
\put (  0,  0) {\line(0,1){280}}
\put (230,  0) {\line(0,1){280}}
\put (310,  0) {\line(0,1){280}}
\thicklines
\put ( 30, 30) {\line(1,0){20}}
\put ( 30, 40) {\line(1,0){20}}
\put ( 30, 60) {\line(1,0){20}}
\put ( 30, 90) {\line(1,0){20}}
\put ( 30,130) {\line(1,0){20}}
\put ( 30,180) {\line(1,0){20}}
\put ( 30,240) {\line(1,0){20}}

\put (130, 30) {\line(1,0){20}}
\put (130, 40) {\line(1,0){20}}
\put (130, 60) {\line(1,0){20}}
\put (130, 90) {\line(1,0){20}}
\put (130,130) {\line(1,0){20}}
\put (130,180) {\line(1,0){20}}
\put (130,240) {\line(1,0){20}}

\put (250, 30) {\line(1,0){20}}
\put (250, 90) {\line(1,0){20}}
\put (250,130) {\line(1,0){20}}
\put (250,240) {\line(1,0){20}}
\small
\put ( 55, 23) {$0^+_{A_1}$}
\put ( 55, 40) {$1^-_{F_2}$}
\put ( 55, 57) {$2^+_{E F_2}$}
\put ( 55, 87) {$3^-_{A_1 F_2 F_1}$}
\put ( 55,127) {$4^+_{A_1 E F_2 F_1}$}
\put ( 55,177) {$5^-_{E F_2 F_2 F_1}$}
\put ( 55,237) {$6^+_{A_1 A_2 E F_2 F_2 F_1}$}

\put (155, 23) {$0^-_{A_2}$}
\put (155, 40) {$1^+_{F_1}$}
\put (155, 57) {$2^-_{E F_1}$}
\put (155, 87) {$3^+_{A_2 F_1 F_2}$}
\put (155,127) {$4^-_{A_2 E F_1 F_2}$}
\put (155,177) {$5^+_{E F_1 F_1 F_2}$}
\put (155,237) {$6^-_{A_2 A_1 E F_1 F_1 F_2}$}

\put (275, 27) {$0^+_{A_1}$}
\put (275, 87) {$3^-_{A_1}$}
\put (275,127) {$4^+_{A_1}$}
\put (275,237) {$6^{\pm}_{A_1}$}

\end{picture}
\caption{Schematic spectrum of the rotational states of the ground stand vibrational 
band with $(v_1 v_2 v_3)=(000)$. All states have $L=I$ and are labeled by $L^P_t$. 
On the right-hand side we show the symmetric states with $t=A_1$.} 
\label{top}
\end{figure}
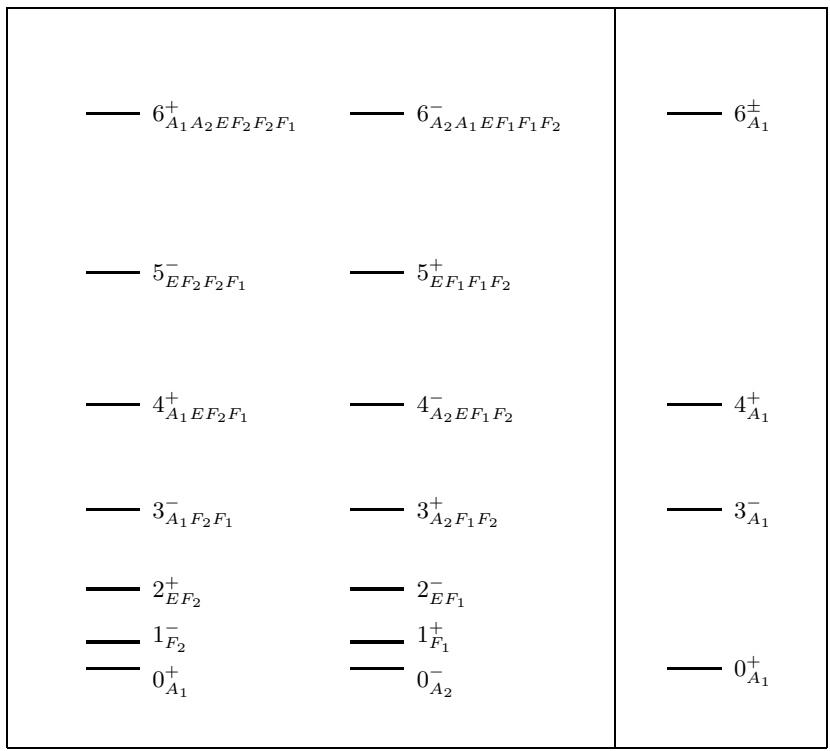

\begin{figure}
\centering
\setlength{\unitlength}{0.7pt} 
\begin{picture}(300,280)(0,0)
\thinlines
\put (  0,  0) {\line(1,0){300}}
\put (  0,280) {\line(1,0){300}}
\put (  0,  0) {\line(0,1){280}}
\put (300,  0) {\line(0,1){280}}
\thicklines
\put ( 30, 60) {\line(1,0){20}}
\put ( 30, 96) {\line(1,0){20}}\put (265,147) {$4^{\pm}$}

\put ( 30,120) {\line(1,0){20}}
\put ( 30,186) {\line(1,0){20}}
\multiput ( 70, 60)(5,0){39}{\circle*{0.1}}
\thinlines
\put ( 25, 25) {$(000)A_1$}
\put ( 55, 57) {$0^+$}
\put ( 55, 93) {$3^-$}
\put ( 55,117) {$4^+$}
\put ( 55,183) {$6^{\pm}$}
\thicklines
\put (100, 90) {\line(1,0){20}}
\put (100,126) {\line(1,0){20}}
\put (100,150) {\line(1,0){20}}
\put (100,216) {\line(1,0){20}}
\multiput (110, 60)(0,5){6}{\circle*{0.1}}
\thinlines
\put ( 95, 25) {$(100)A_1$}
\put (125, 87) {$0^+$}
\put (125,123) {$3^-$}
\put (125,147) {$4^+$}
\put (125,213) {$6^{\pm}$}
\thicklines
\put (170,108) {\line(1,0){20}}
\put (170,150) {\line(1,0){20}}
\put (170,180) {\line(1,0){20}}
\put (170,216) {\line(1,0){20}}
\multiput (180, 60)(0,5){10}{\circle*{0.1}}
\thinlines
\put (165, 25) {$(010)E$}
\put (195,105) {$2^{\pm}$}
\put (195,147) {$4^{\pm}$}
\put (195,177) {$5^{\pm}$}
\put (195,213) {$6^{\pm}$}
\thicklines
\put (240, 96) {\line(1,0){20}}
\put (240,108) {\line(1,0){20}}
\put (240,126) {\line(1,0){20}}
\put (240,150) {\line(1,0){20}}
\put (240,180) {\line(1,0){20}}
\put (240,216) {\line(1,0){20}}
\multiput (250, 60)(0,5){8}{\circle*{0.1}}
\thinlines
\put (235, 25) {$(001)F_2$}
\put (265, 93) {$1^-$}
\put (265,105) {$2^+$}
\put (265,123) {$3^{\pm}$}
\put (265,147) {$4^{\pm}$}
\put (265,177) {$5^{-\pm}$}
\put (265,213) {$6^{+\pm}$}
\thicklines
\put(20,220) {\circle*{10}} 
\put(60,220) {\circle*{10}}
\put(70,240) {\circle*{10}}
\put(40,260) {\circle*{10}}
\put(20,220) {\line( 1,0){40}}
\put(20,220) {\line( 1,2){20}}
\put(60,220) {\line(-1,2){20}}
\put(60,220) {\line( 1,2){10}}
\put(70,240) {\line(-3,2){30}}
\multiput(20,220)(5,2){11}{\circle*{2}}
\end{picture}
\caption{Schematic spectrum of a spherical top with tetrahedral symmetry 
and $\omega_1=\omega_2=\omega_3$. The rotational bands are labeled by 
$(v_1 v_2 v_3)$ (bottom). All states are symmetric under $S_4 \sim {\cal T}_d$.} 
\label{sphtop}
\end{figure}
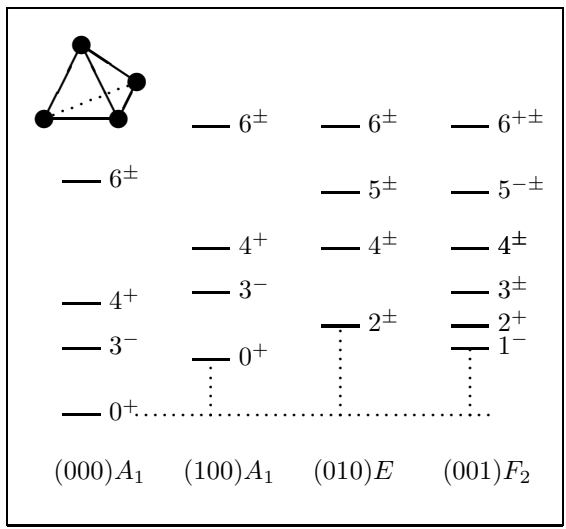

Fig.~\ref{top} shows the structure of the rotational excitations of the ground 
state band $(v_1 v_2 v_3)=(000)$. The rotational levels are doubled because of 
inversion doubling: for each value of the angular momentum $L$, one has doublets of 
states with ($A_1$, $A_2$), ($E$, $E$) and ($F_2$, $F_1$), in agreement with the 
classification of rotational levels of a spherical top with tetrahedral symmetry 
\cite{Herzberg,Oka}. For identical bosons, as is the case for a cluster of four 
$\alpha$-particles, the allowed rotational-vibrational states are the symmetric 
ones with $t=A_1$, and therefore the states of the ground state band have angular 
momentum and parity $L^P=0^+$, $3^-$, $4^+$, $6^{\pm}$, $\ldots$, as shown in the 
right-hand side of Fig.~\ref{top}. 
A similar analysis can be done for the rotational bands built on the $(100)A_1$, 
$(010)E$ and $(001)F_2$ vibrations. For the $A_1$ vibration the values of angular momentum 
and parity are the same as for the ground state band  $L^P=0^+$, $3^-$, $4^+$, 
$6^{\pm}$, $\ldots$. For the doubly degenerate $E$ vibration they are $L^P=2^{\pm}$, 
$4^{\pm}$, $5^{\pm}$, $6^{\pm}$, $\ldots$, while for the triply degenerate $F_2$ vibration 
they are  $L^P=1^-$, $2^+$, $3^{\pm}$, $4^{\pm}$, $5^{-,\pm}$, $6^{+,\pm}$ $\ldots$. 
The situation is summarized in Fig.~\ref{sphtop} which shows the expected spectrum 
of a spherical top with tetrahedral symmetry and $\omega_1=\omega_2=\omega_3$. 

\section{Transition probabilities}

In order to calculate transition form factors and transition probabilities 
in the algebraic cluster model one has to express the transition operators 
in terms of the algebraic operators. The transition form factors are the 
matrix elements of $\sum_{i=1}^4 \exp(i \vec{q} \cdot \vec{r}_i)$ where 
$\vec{q}$ is the momentum transfer. Choosing the $z$-axis along the direction 
of the momentum transfer and using the fact that the four particles are 
identical, it is sufficient to consider the matrix elements of $\exp(i q r_{4z})$. 
After converting to Jacobi coordinates and integrating over the center-of-mass 
coordinate one has $\exp(-i q \sqrt{3/4} \, \eta_{z})$. The matrix elements of 
this operator can be obtained algebraically by making the replacement  
\ba
\sqrt{3/4} \, \eta_z \;\rightarrow\; \beta \hat D_{\eta,z}/X_D ~, 
\label{map} 
\ea
where $\beta$ represents the scale of the coordinate and $X_D$ is given 
by the reduced matrix element of the dipole operator. 
The replacement in Eq.~(\ref{map}) comes from the fact that in the large $N$ 
limit, the dipole operators $\hat{D}_{\rho}$, $\hat{D}_{\lambda}$ and 
$\hat{D}_{\eta}$ of Eq.~(\ref{gen}) correspond to the three Jacobi 
coordinates $\vec{\rho}$, $\vec{\lambda}$ and $\vec{\eta}$ \cite{onno}.
 
In summary, the transition form factors can be expressed in the ACM as 
\ba
{\cal F}_M(i \rightarrow f;q) \;=\; \langle \gamma_f, L_f, M \, 
| \, \hat T(\epsilon) \, | \, \gamma_i, L_i, M \rangle ~, 
\ea
with
\ba
\hat T(\epsilon) \;=\; e^{i \epsilon \hat{D}_{\eta,z}} \;=\; e^{-iq\beta \hat{D}_{\eta,z}/X_{D}} ~. 
\label{trans}
\ea
The transition probabilities $B(EL)$ can be extracted from the form factors 
in the long wavelength limit
\ba
B(EL;i \rightarrow f) \;=\; (Ze)^2 \, \frac{[(2L+1)!!]^{2}}{4\pi (2L_{i}+1)} 
\, \lim_{q\rightarrow 0} \sum_M \frac{
\left| {\cal F}_M(i \rightarrow f;q) \right|^{2}}{q^{2L}} ~,
\label{belif}
\ea
where $Ze$ is the total electric charge of the cluster. 
In the three special cases of the harmonic oscillator, or $U(9)$ limit, 
the deformed oscillator, or $SO(10)$ limit, and the spherical top with 
${\cal T}_d$ symmetry, the matrix elements of $\hat T$ can be obtained 
analytically and are given in the subsections below. 

For the general ACM Hamiltonian of Eq.~(\ref{HS4}), 
the matrix elements cannot be obtained in closed analytic form, but have to 
be calculated numerically. Hereto, a computer program has been developed 
\cite{ffacm}, in which the form factors are obtained exactly by using 
the symmetry properties of the transition operator of Eq.~(\ref{trans}). 

When a dynamic symmetry occurs the matrix elements of the transition 
operators can be obtained in explicit analytic form. The general procedure 
to derive the transition form factors in the harmonic oscillator  
and deformed oscillator limits was discussed in \cite{BG,Manko}. 

In the $U(9)$ limit of the ACM the elastic form factor can be derived as  
\ba
{\cal F}(0^+ \rightarrow 0^+;q) \;=\; (\cos \epsilon)^N   
\;\rightarrow\; e^{-q^2 \beta^2/6} ~,
\label{ffel1}
\ea
with $\epsilon=-q\beta/X_D$ and $X_D=\sqrt{3N}$. In the large $N$ limit, the elastic form 
factor exhibits an exponential fall-off with momentum transfer.   

In the $SO(10)$ limit the elastic form factor is given in terms of a Gegenbauer polynomial
\ba
{\cal F}(0^+ \rightarrow 0^+;q) \;=\; \frac{7!N!}{(N+7)!} \, C^{(4)}_N(\cos \epsilon) 
\;\rightarrow\; \frac{7!! \, j_3(q\beta \sqrt{3})}{(q \beta \sqrt{3})^3} ~. 
\label{ffel2}
\ea
with $\epsilon=-q\beta/X_D$ and $X_D=\sqrt{N(N+8)/3}$. In the large $N$ limit, the elastic form 
factor is proportional to a spherical Bessel function. 

For the spherical top with ${\cal T}_d$ symmetry, the form factors 
can only be obtained in closed form in the large $N$ limit using a technique 
introduced in \cite{BAS} and subsequently exploited in \cite{BIL1}. The normalization 
factor $X_D$ which appears in the algebraic transition operator of Eq.~(\ref{trans}), 
is here $X_D=2NR/(1+R^2)\sqrt{3}$. 
The elastic form factor can be obtained as  
\ba
{\cal F}(0^+ \rightarrow 0^+;q) &\rightarrow& \frac{1}{4\pi} \int d \Omega 
\left< N,c \left| {\cal R}^{-1}(\Omega) \hat{T}(\epsilon) {\cal R}(\Omega) \right| N,c \right> 
\nonumber\\ 
&=& \frac{1}{2} \int d \cos \theta \, e^{-i q \beta \cos \theta} = j_0(q \beta) ~.  
\label{ffel3}
\ea
In Eq.~(\ref{ffel3}), $|N,c\rangle$ is the condensate corresponding to a tetrahedral equilibrium configuration
\ba
\left| N,c \right> &=& \frac{1}{\sqrt{N!}} \left( b_c^{\dagger} \right)^N \left| 0 \right> ~,
\nonumber\\ 
b_c^{\dagger} &=& \frac{1}{\sqrt{1+R^2}} \left( s^{\dagger} + \frac{R}{\sqrt{3}}   
\left( b^{\dagger}_{\rho x} + b^{\dagger}_{\lambda y} + b^{\dagger}_{\eta z} \right) \right) ~.
\ea
In general, for transitions along the ground state band $(000)A_1$ the transition 
form factors are  given in terms of a spherical Bessel function  
\ba
{\cal F}(0^+ \rightarrow L^P;q) \;\rightarrow\; c_L \, j_L(q \beta) ~, 
\ea
with
\ba
c_L^2 \;=\; \frac{2L+1}{16} \left[ 4+12P_{L}(-\frac{1}{3}) \right] ~. 
\ea
The coefficients $c_1^2$, $c_2^2$ and $c_5^2$ vanish as a consequence of the tetrahedral 
symmetry. Some values which are relevant to the lowest states are $c_0^2=1$, $c_3^2=35/9$, 
$c_4^2=7/3$ and $c_6^2=416/81$.

\begin{table}
\centering
\caption[Spherical top form factors]
{Transition form factors for the ground state band of the spherical top.}
\label{fftop}
\vspace{5pt}
\begin{tabular}{cc}
\hline
\noalign{\smallskip}
$L^P$ & $\left| {\cal F}(0^+ \rightarrow L^P;q) \right|$ \\
\noalign{\smallskip}
\hline
\noalign{\smallskip}
$0^+$ & $j_0(q\beta)$ \\
\noalign{\smallskip}
$3^-$ & $\sqrt{\frac{35}{9}} \, j_3(q\beta)$ \\
\noalign{\smallskip}
$4^+$ & $\sqrt{\frac{7}{3}} \, j_4(q\beta)$ \\
\noalign{\smallskip}
$6^+$ & $\sqrt{\frac{416}{81}} \, j_6(q\beta)$ \\
\noalign{\smallskip}
\hline 
\end{tabular}
\end{table}

The transition probabilities $B(EL)$ along the ground state band can be extracted 
from the form factors in the long wavelength limit according to Eq.~(\ref{belif}). 
The result is given by a simple formula
\ba
B(EL;0^+ \rightarrow L^P) \;=\; \left(\frac{Ze\beta^{L}}{4}\right)^{2} 
\frac{2L+1}{4\pi} \left[ 4+12P_{L}(-\frac{1}{3}) \right] ~, 
\label{BEL}
\ea
with $P=(-1)^L$. Explicit expressions are
\ba
B(E3;0^+ \rightarrow 3^-) &=& (Ze)^{2} \frac{7}{4\pi} \frac{5}{9} \beta^{6} ~,
\nonumber \\
B(E4;0^+ \rightarrow 4^+) &=& (Ze)^{2} \frac{9}{4\pi} \frac{7}{27} \beta^{8} ~,
\nonumber \\
B(E6;0^+ \rightarrow 6^+) &=& (Ze)^{2} \frac{13}{4\pi} \frac{32}{81}\beta^{12} ~.
\ea
Form factors and $B(EL)$ values only depend on the parameter $\beta$, the distance 
of each particle from the center of mass of the tetrahedral configuration, and on 
the ${\cal T}_d$ symmetry which gives the coefficients $c_L$. The analytic results 
given in this section provide a set of closed expressions which can be compared 
with experiment.

All results for form factors given in Eqs.~(\ref{ffel1}-\ref{ffel3}) and Table~\ref{fftop} 
are for point-like constituent particles with a charge distribution 
\ba
\rho (\vec{r}) \;=\; \frac{Ze}{4} \sum_{i=1}^{4} \delta (\vec{r}-\vec{r}_{i}) ~,
\label{charge}
\ea
where $Ze$ is the total electric charge. In cases in which the constituent particles 
are composite particles and thus have an intrinsic form factor they must be modified. 
Assuming a Gaussian form, 
\ba
\rho (\vec{r}) \;=\; \frac{Ze}{4} \left( \frac{\alpha}{\pi} \right)^{3/2} 
\sum_{i=1}^{4} \exp \left[ -\alpha \left( \vec{r}-\vec{r}_{i} \right)^{2}\right] ~, 
\label{rhor1}
\ea
all form factors are multiplied by an exponential factor $\exp(-q^{2}/4\alpha)$. 
The charge radius can be obtained from the slope of the elastic form 
factor in the origin 
\ba
\langle r^{2} \rangle^{1/2} \;=\; \left[ -6 \left. 
\frac{d {\cal F}(0^{+} \rightarrow 0^{+};q)}{dq^{2}} \right|_{q=0} \right]^{1/2} 
\;=\; \sqrt{\frac{3}{2\alpha}+\beta^{2}} ~. 
\label{radius}
\ea
The above formula for the charge radius is valid for all three cases discussed in this 
section: the harmonic oscillator, the deformed oscillator and the spherical top. 

It is of great interest to determine the charge distribution in the case in which one has 
composite particles at the vertices of a tetrahedron. This charge distribution is given by 
the expansion of Eq.~(\ref{rhor1}) into multipoles
\ba
\rho (\vec{r}) \;=\; \frac{Ze}{4} \left( \frac{\alpha}{\pi} \right)^{3/2} e^{-\alpha(r^{2}+\beta^{2})}  
\sum_{\lambda=0}^{\infty} (2\lambda+1) \, i_{\lambda}(2\alpha \beta r) 
\sum_{i=1}^{4} P_{\lambda}(\cos \gamma_i) ~,
\label{rhor}
\ea
with
\ba
\cos \gamma_i \;=\; \cos \theta \cos \theta_i + \sin \theta \sin \theta_i \cos(\phi-\phi_i) ~.
\ea
Here we have used spherical coordinates $\vec{r}=(r,\theta,\phi)$ and $\vec{r}_i=(\beta,\theta_i,\phi_i)$.  
The angles $\theta_i$ and $\phi_i$ denote the location of the four particles, and 
$i_{\lambda}(x)=j_{\lambda}(ix)/i^{\lambda}$ is the modified spherical Bessel function. 
This formula is valid for all cases in which the distance from the origin is the same $|\vec{r}_i|=\beta$ 
for all constituents. In the case of $n$ constituents, the charge of the constituent 
in Eq.~(\ref{rhor}) is changed from $Ze/4$ to $Ze/n$, and the sum is over all constituents from $1$ to $n$.

\section{The nucleus $^{16}$O}

The nucleus $^{16}$O has been the subject of many investigations. Within the framework of the  
nuclear shell model detailed studies were made by Zuker {\em et al.} \cite{Zuker} in terms 
of 1p-1h and by Brown and Green \cite{Brown} and Feshbach and Iachello \cite{Feshbach} in terms 
of multiparticle-multihole configurations. However, as early as 1954, Dennison suggested that its 
spectrum could be understood in terms of an $\alpha$-particle model with ${\cal T}_d$ symmetry 
\cite{Dennison}. This idea was adopted by Kameny \cite{Kameny}, Brink \cite{Brink1,Brink2}, and 
especially by Robson \cite{Robson1,Review,Robson2,Robson3}, who in a series of papers developed 
the model in further detail. Other possible configurations were also studied, in particular 
$\alpha + ^{12}$C \cite{Suzuki}. 

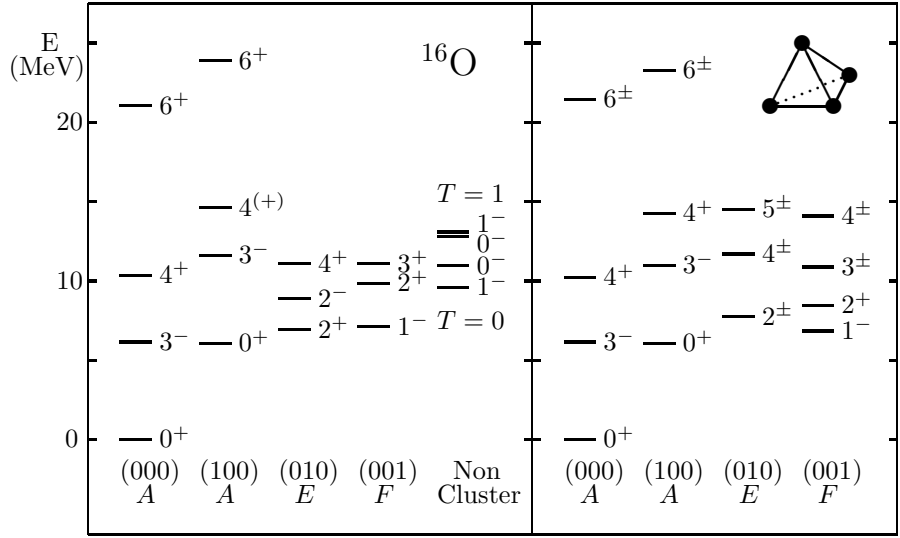
\begin{figure}
\centering
\vspace{15pt}
\setlength{\unitlength}{0.6pt}
\begin{picture}(560,335)(-50,0)
\normalsize
\thinlines
\put (  0,  0) {\line(0,1){335}}
\put (  0,  0) {\line(1,0){510}}
\put (  0,335) {\line(1,0){510}}
\put (280,  0) {\line(0,1){335}}
\put (510,  0) {\line(0,1){335}}
\thicklines
\put (  0, 60) {\line(1,0){5}}
\put (  0,110) {\line(1,0){5}}
\put (  0,160) {\line(1,0){5}}
\put (  0,210) {\line(1,0){5}}
\put (  0,260) {\line(1,0){5}}
\put (  0,310) {\line(1,0){5}}
\put (275, 60) {\line(1,0){10}}
\put (275,110) {\line(1,0){10}}
\put (275,160) {\line(1,0){10}}
\put (275,210) {\line(1,0){10}}
\put (275,260) {\line(1,0){10}}
\put (275,310) {\line(1,0){10}}
\put (505, 60) {\line(1,0){5}}
\put (505,110) {\line(1,0){5}}
\put (505,160) {\line(1,0){5}}
\put (505,210) {\line(1,0){5}}
\put (505,260) {\line(1,0){5}}
\put (505,310) {\line(1,0){5}}
\put (-20, 55) { 0}
\put (-20,155) {10}
\put (-20,255) {20}
\put (-30,305) {E}
\put (-50,290) {(MeV)}
\put ( 20, 60.00) {\line(1,0){20}}
\put ( 20,121.30) {\line(1,0){20}}
\put ( 20,163.56) {\line(1,0){20}}
\put ( 20,270.52) {\line(1,0){20}}
\put ( 20, 35) {$(000)$}
\put ( 30, 20) {$A$}
\put ( 45, 55.00) {$0^+$}
\put ( 45,116.30) {$3^-$}
\put ( 45,158.56) {$4^+$}
\put ( 45,265.52) {$6^+$}
\put ( 70,120.49) {\line(1,0){20}}
\put ( 70,176.00) {\line(1,0){20}}
\put ( 70,206.20) {\line(1,0){20}}
\put ( 70,298.79) {\line(1,0){20}}
\put ( 70, 35) {$(100)$}
\put ( 80, 20) {$A$}
\put ( 95,115.49) {$0^+$}
\put ( 95,171.00) {$3^-$}
\put ( 95,201.20) {$4^{(+)}$}
\put ( 95,293.79) {$6^+$}
\put (120,129.17) {\line(1,0){20}}
\put (120,148.72) {\line(1,0){20}}
\put (120,170.97) {\line(1,0){20}}
\put (120, 35) {$(010)$}
\put (130, 20) {$E$}
\put (145,124.17) {$2^+$}
\put (145,143.72) {$2^-$}
\put (145,165.97) {$4^+$}
\put (170,131.16) {\line(1,0){20}}
\put (170,158.44) {\line(1,0){20}}
\put (170,170.80) {\line(1,0){20}}
\put (170, 35) {$(001)$}
\put (180, 20) {$F$}
\put (195,126.16) {$1^-$}
\put (195,153.44) {$2^+$}
\put (195,165.80) {$3^+$}
\put (220,155.85) {\line(1,0){20}}
\put (220,169.57) {\line(1,0){20}}
\put (220,187.96) {\line(1,0){20}}
\put (220,190.90) {\line(1,0){20}}
\put (230, 35) {Non}
\put (220, 20) {Cluster}
\put (245,150.85) {$1^-$}
\put (245,164.57) {$0^-$}
\put (220,130) {$T=0$}
\put (245,177.96) {$0^-$}
\put (245,190.90) {$1^-$}
\put (220,210) {$T=1$}
\Large
\put (210,290) {$^{16}$O}
\normalsize
\put (300, 60.0) {\line(1,0){20}}
\put (300,121.3) {\line(1,0){20}}
\put (300,162.2) {\line(1,0){20}}
\put (300,274.6) {\line(1,0){20}}
\put (300, 35) {$(000)$}
\put (310, 20) {$A$}
\put (325, 55.00) {$0^+$}
\put (325,116.30) {$3^-$}
\put (325,157.2) {$4^+$}
\put (325,269.6) {$6^{\pm}$}
\put (350,120.5) {\line(1,0){20}}
\put (350,169.7) {\line(1,0){20}}
\put (350,202.5) {\line(1,0){20}}
\put (350,292.7) {\line(1,0){20}}
\put (350, 35) {$(100)$}
\put (360, 20) {$A$}
\put (375,115.5) {$0^+$}
\put (375,164.7) {$3^-$}
\put (375,197.5) {$4^+$}
\put (375,287.7) {$6^{\pm}$}
\put (400,137.4) {\line(1,0){20}}
\put (400,176.9) {\line(1,0){20}}
\put (400,205.1) {\line(1,0){20}}
\put (400, 35) {$(010)$}
\put (410, 20) {$E$}
\put (425,132.4) {$2^{\pm}$}
\put (425,171.9) {$4^{\pm}$}
\put (425,200.1) {$5^{\pm}$}
\put (450,128.5) {\line(1,0){20}}
\put (450,144.6) {\line(1,0){20}}
\put (450,168.7) {\line(1,0){20}}
\put (450,200.9) {\line(1,0){20}} 
\put (450, 35) {$(001)$}
\put (460, 20) {$F$}
\put (475,123.5) {$1^-$}
\put (475,139.6) {$2^+$}
\put (475,163.7) {$3^{\pm}$}
\put (475,195.9) {$4^{\pm}$}
\put(430,270) {\circle*{10}} 
\put(470,270) {\circle*{10}}
\put(480,290) {\circle*{10}}
\put(450,310) {\circle*{10}}
\put(430,270) {\line( 1,0){40}}
\put(430,270) {\line( 1,2){20}}
\put(470,270) {\line(-1,2){20}}
\put(470,270) {\line( 1,2){10}}
\put(480,290) {\line(-3,2){30}}
\multiput(430,270)(5,2){11}{\circle*{2}}
\end{picture}
\caption{Comparison between the observed spectrum of $^{16}$O (left) and the theoretical 
spectrum (right). The levels are organized in columns corresponding to the ground state 
band and the three vibrational bands with $A$, $E$ and $F$ symmetry of a spherical top 
with tetrahedral symmetry. The last column shows the lowest non-cluste r levels.} 
\label{O16}
\end{figure}

In very recent years, $^{16}$O has been again the subject of many investigations, both within the 
framework of the no-core shell model \cite{Navratil} and {\em ab initio} lattice calculations 
\cite{lattice}. The latter shows evidence for tetrahedral symmetry in the ground state of $^{16}$O. 
Another important development has been the suggestion that some of the excited $0^+$ states may 
have a large $\alpha$-condensate fraction and/or $\alpha + ^{12}$C cluster structure 
\cite{Funaki1,Funaki2}. 

In this article, we analyze the available data in the terms of the spherical top limit of the $U(10)$ 
algebraic cluster model with ${\cal T}_d$ symmetry.  

\begin{table}[t]
\small
\centering
\caption{Classification of levels in $^{16}$O in terms of ${\cal T}_d$ symmetry for 
the ground state band and single vibrational excitations.}
\label{levels1}
\vspace{5pt}
\begin{tabular}{cccc}
\hline 
\noalign{\smallskip}
$(v_1,v_2,v_3)$ & $L^P$ & $E_{\rm th}$(MeV) & $E_{\rm exp}$(MeV) \\
\noalign{\smallskip}
\hline
\noalign{\smallskip}
$(000)A_1$ & $0^+$ &  0.00 &  0.000 \\
            & $3^-$ &  6.13 &  6.130 \\
            & $4^+$ & 10.22 & 10.356 \\
            & $6^+$ & 21.46 & 21.052 \\
            & $6^-$ & 21.46 &        \\
            & $7^-$ & 28.62 & (28.2) \\
            & $8^+$ & 36.79 &        \\
\noalign{\smallskip}
\hline 
\noalign{\smallskip}
$(100)A_1$ & $0^+$ &  6.05 &  6.049 \\
            & $3^-$ & 10.97 & 11.600 \\
            & $4^+$ & 14.25 & 14.620 \\
            & $6^+$ & 23.27 & 23.880 \\
            & $6^-$ & 23.27 &        \\
            & $7^-$ & 29.01 & (29.0) \\
            & $8^+$ & 35.57 &        \\
\noalign{\smallskip}
\hline 
\noalign{\smallskip}
$(010)E$ & $2^+$ &  7.74 &  6.917 \\
            & $2^-$ &  7.74 &  8.872 \\
            & $4^+$ & 11.69 & 11.097 \\
            & $4^-$ & 11.69 &        \\
            & $5^+$ & 14.51 & 14.399 \\
            & $5^-$ & 14.51 & 14.660 \\
            & $6^+$ & 17.89 & (16.275) \\
            & $6^-$ & 17.89 &        \\
\noalign{\smallskip}
\hline 
\noalign{\smallskip}
$(001)F_2$ & $1^-$ &  6.85 &  7.117 \\
            & $2^+$ &  8.46 &  9.844 \\
            & $3^+$ & 10.87 & 11.080 \\
            & $3^-$ & 10.87 &        \\
            & $4^+$ & 14.09 & (13.869) \\
            & $4^-$ & 14.09 & (14.302) \\
            & $5^{\pm}$ & 18.11 &  \\
            & $5^-$     & 18.11 &  \\
            & $6^{\pm}$ & 22.93 &  \\
            & $6^+$     & 22.93 & (23.0) \\
\noalign{\smallskip}
\hline 
\end{tabular}
\end{table}

\subsection{Energies}

There are 340 levels known in $^{16}$O in the energy range $0-35$ MeV \cite{Tilley}. We have first 
analyzed these levels with the simple formula for a rigid spherical top with tetrahedral symmetry 
\ba 
E(v_1,v_2,v_3,L) \;=\; E_0 + \omega_{1} \, v_{1} + \omega_{2} \, v_{2} + \omega_{3} \, v_{3} 
+ B_{[v]} L(L+1) ~,
\label{sphericaltop}
\ea
where the zero-point energy is given by $E_0= \frac{1}{2} \omega_{1}+ \omega_{2} + \frac{3}{2} \omega_{3}$. 
By taking $\omega_1=\omega_2=\omega_3=6.05$ MeV, $B_{000A}=0.511$ MeV, $B_{100A}=0.410$ MeV, 
$B_{010E}=0.282$ MeV and $B_{001F}=0.402$ MeV, we are able to obtain an excellent description 
of the low-lying spectrum of $^{16}$O, a portion of which is shown in Fig.~\ref{O16}. Our classification 
includes 17 states firmly assigned to ${\cal T}_d$ symmetry and another 11 tentatively assigned (in parentheses) 
covering the range from 0 to 29 MeV, as shown in Tables~\ref{levels1} and \ref{levels2}. 
We note that because of the multiplication rule 
$E \otimes F_2 = F_1 \oplus F_2$ the vibration $(011)$ contains $F_1$.  

\begin{figure}
\centering
\includegraphics[width=4in]{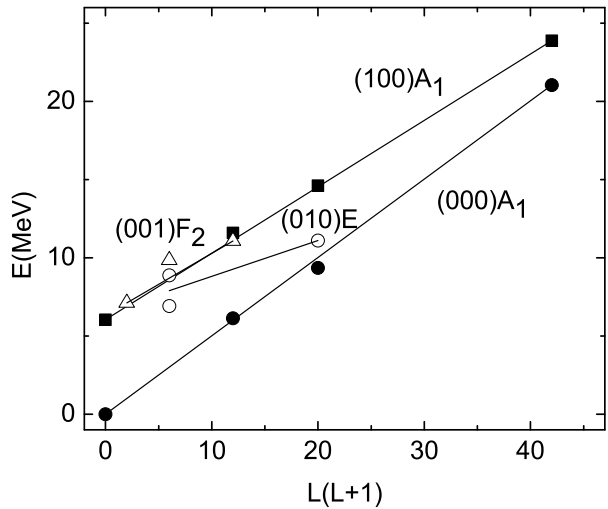}
\vspace{-0.5cm}
\caption[]{The excitation energies of cluster states in $^{16}$O 
plotted as a function of $L(L+1)$: closed circles for the ground state band 
($0^+$, $3^-$, $4^+$, $6^+$), closed squares for the $A_1$ vibration 
($0^+$, $3^-$, $4^+$, $6^+$), open circles for the $E$ vibration ($2^+$, $2^-$, $4^+$)
and open triangles for the $F_2$ vibration ($1^-$, $2^+$, $3^+$) \cite{TdO16}.} 
\label{bands}
\end{figure}

The rotational ground state band with angular momenta $L^P=0^+$, $3^-$, $4^+$, $6^+$ has been 
observed with moment of inertia such that $B_{000A}=0.511$ MeV. The spherical top predicts a 
$L^P=6^{\pm}$ doublet as a consequence of the tetrahedral symmetry. The $6^-$ state has not been 
identified yet. It appears that all three vibrations $(100)A_1$, $(010)E$ and $(001)F_2$ have been 
identified with comparable energy $\omega_1=\omega_2=\omega_3=6.05$ MeV as one would expect from 
Eq.~(\ref{h3vib}) with $\xi_1=\xi_2=\xi_3$. Members of the rotational bands have also been observed. 
With the present assignments of $L^P=0^+$, $3^-$, $4^+$, $6^+$ states of the the breathing mode 
$(100)A$, this vibrational band has a moment of inertia very similar to that of the ground state band. 
This band is similar in nature to the band built on the Hoyle state in $^{12}$C which was recently 
measured experimentally \cite{Marin,Itoh,Freer2,Gai,Freer4} and reviewed in \cite{Hoyle}. 
The moments of inertia of the $(100)A_1$, $(010)E$ and $(001)F_2$ bands are larger (smaller $B$ values) 
than that of the ground state band due to their nature (breathing and bending vibrations). 
The moments of inertia of the $A_1$ and $F_2$ vibrations have almost the same value (and somewhat 
larger than that of the ground state band), whereas the moment of inertia of the $E$ vibration has 
almost double the value of that of the ground state band. The situation is summarized in Fig.~\ref{bands}. 
It appears also that some members of the double vibrations $(200)A_1$, $(110)E$, $(101)F_2$, 
$(002)A_1$ and $(011)F_1$ can be identified. The vibrations are nearly harmonic. 

A characteristic feature of the spectrum of a spherical top with tetrahedral symmetry is the 
occurrence of parity doublets both in the ground state band and in the vibrational excitations 
(see Fig.~\ref{sphtop}). In the ground state band and the $(100)A_1$ breathing vibration one 
expects an additional $L^P=6^-$ state, in the $(010)E$ vibration an extra $4^-$ state and in 
the $(001)F_2$ vibration a $3^-$ state is missing (see Table~\ref{levels1}).  

\begin{table}
\small
\centering
\caption[Classification]
{Classification of levels in $^{16}$O in terms of ${\cal T}_d$ symmetry for 
double vibrational excitations.}
\label{levels2}
\vspace{5pt}
\begin{tabular}{cccc}
\hline
\noalign{\smallskip}
$(v_1,v_2,v_3)$ & $L^P$ & $E_{\rm th}$(MeV) & $E_{\rm exp}$(MeV) \\
\noalign{\smallskip}
\hline
\noalign{\smallskip}
$(200)A_1$ & $0^+$ & 12.10 & 12.049 \\
            & $3^-$ & &  \\
\noalign{\smallskip}
\hline
\noalign{\smallskip}
$(110)E$ & $2^+$ & & (11.520) \\
            & $2^-$ & &  \\
\noalign{\smallskip}
\hline
\noalign{\smallskip}
$(101)F_2$ & $1^-$ & & (12.440) \\
            & $2^+$ & & (13.020) \\
\noalign{\smallskip}
\hline
\noalign{\smallskip}
$(020)A_1$ & $0^+$ & 12.10 & \\
$(020)E$ & $2^-$ & & \\
            & $2^+$ & & \\
\noalign{\smallskip}
\hline
\noalign{\smallskip}
$(002)A_1$ & $0^+$ & 12.10 & (11.260) \\
\noalign{\smallskip}
\hline
\noalign{\smallskip}
$(011)F_1$ & $1^+$ & & (13.664) \\
            & $1^-$ & & \\
\noalign{\smallskip}
\hline 
\end{tabular}
\end{table}

The experimental spectrum is very similar to that of a spherical top with tetrahedral symmetry. 
Nevertheless there are some perturbations which can be described by adding higher order terms 
to the Hamiltonian, and consequently to the energy formula of Eq.~(\ref{sphericaltop}). 
Simple perturbations are\\
(i) vibrational anharmonicities which contribute as $x_{ij}v_iv_j$, \\
(ii) centrifugal stretching which leads to corrections of the form $D_{[v]} L^2(L+1)^2$ \\
(iii) Coriolis coupling. This only affects $F$ representations and causes a splitting of the 
triply degenerate vibration into three pieces
\ba
F^{(+)} &=& B_{[v]} L(L+1) + 2B_{[v]} \zeta (L+1) ~,
\nonumber\\
F^{(0)} &=& B_{[v]} L(L+1) ~,
\nonumber\\
F^{(-)} &=& B_{[v]} L(L+1) - 2B_{[v]} \zeta (L+1) ~.
\ea 
According to an estimate by Dennison, the coefficient $\zeta=1/2$ in $^{16}$O \cite{Denn}.\\
(iv) Rotation-vibration interaction. The most notable consequence is a signature splitting 
\ba
\Delta \;=\; \left[ \frac{1+(-1)^P}{2} \right] \eta_{[v]} ~,
\ea
which splits the positive parity from the negative parity states with the same angular momentum, 
for example $2^+$ and $2^-$. This effect is particularly important for the $E$ vibration. 

However, deviations from the simple energy formula of Eq.~(\ref{sphericaltop}) appear to be small 
with the only exception of the $2^+/2^-$ splitting in the $E$ vibration which corresponds to 
$\Delta=1.955$ MeV, and we therefore do not discuss higher order terms further. 
Our assignments are in part different from those of Robson \cite{Review} who introduced a large 
centrifugal stretching and other perturbations. Within the framework of the ACM, perturbations 
can be studied by diagonalizing the Hamiltonian for finite $N$ and $R^2 \neq 1$. 

Our discussion of the spectrum of $^{16}$O has been so far assuming no internal excitation of the 
$\alpha$-particle. However, in addition to the collective cluster excitations we expect also 
single-particle (shell model) excitations. We can identify uniquely non-cluster states by their 
isotopic spin nature ($T=1$) and for $T=0$ by the fact that some spin-parity states are not allowed by
the symmetry of the $4\alpha$ configuration. Specifically, no $0^-$ state can be formed in states 
with $v=v_1+v_2+v_3<3$. We are therefore able to identify non-cluster states in the low-lying 
spectrum as shown in Table~\ref{noncluster}. Some of the non-cluster states are shown in Fig.~\ref{O16}, 
to emphasize the fact that all observed states below 9.58 MeV are cluster states. In a shell-model 
language, cluster states are multiparticle-multihole excitations \cite{Brown,Feshbach}. 

\begin{table}
\small
\centering
\caption[Classification]
{Classification of levels in $^{16}$O in terms of specific shell-model configurations.} 
\label{noncluster}
\vspace{5pt}
\begin{tabular}{cccr}
\hline
\noalign{\smallskip}
Classification & $L^P$ & $E_{\rm exp}$(MeV) \\
\noalign{\smallskip}
\hline
\noalign{\smallskip}
1p-1h $p_{1/2}^{-1} s_{1/2}$ & $1^-$ $T=0$ &  9.585 \\
                             & $0^-$ $T=0$ & 10.957 \\
\noalign{\smallskip}
\hline
\noalign{\smallskip}
1p-1h $p_{1/2}^{-1} d_{5/2}$ & $2^-$ $T=0$ & 12.530 \\
                             & $3^-$ $T=0$ & 13.129 \\
\noalign{\smallskip}
\hline
\noalign{\smallskip}
1p-1h $p_{1/2}^{-1} s_{1/2}$ & $0^-$ $T=1$ & 12.796 \\
                             & $1^-$ $T=1$ & 13.090 \\
\noalign{\smallskip}
\hline
\noalign{\smallskip}
1p-1h $p_{1/2}^{-1} d_{5/2}$ & $2^-$ $T=1$ & 12.968 \\
                             & $3^-$ $T=1$ & 13.259 \\
\noalign{\smallskip}
\hline 
\end{tabular}
\end{table}

\subsection{Form factors and electromagnetic transition rates}

\begin{figure}

\begin{minipage}{.5\linewidth}
\centerline{\epsfig{file=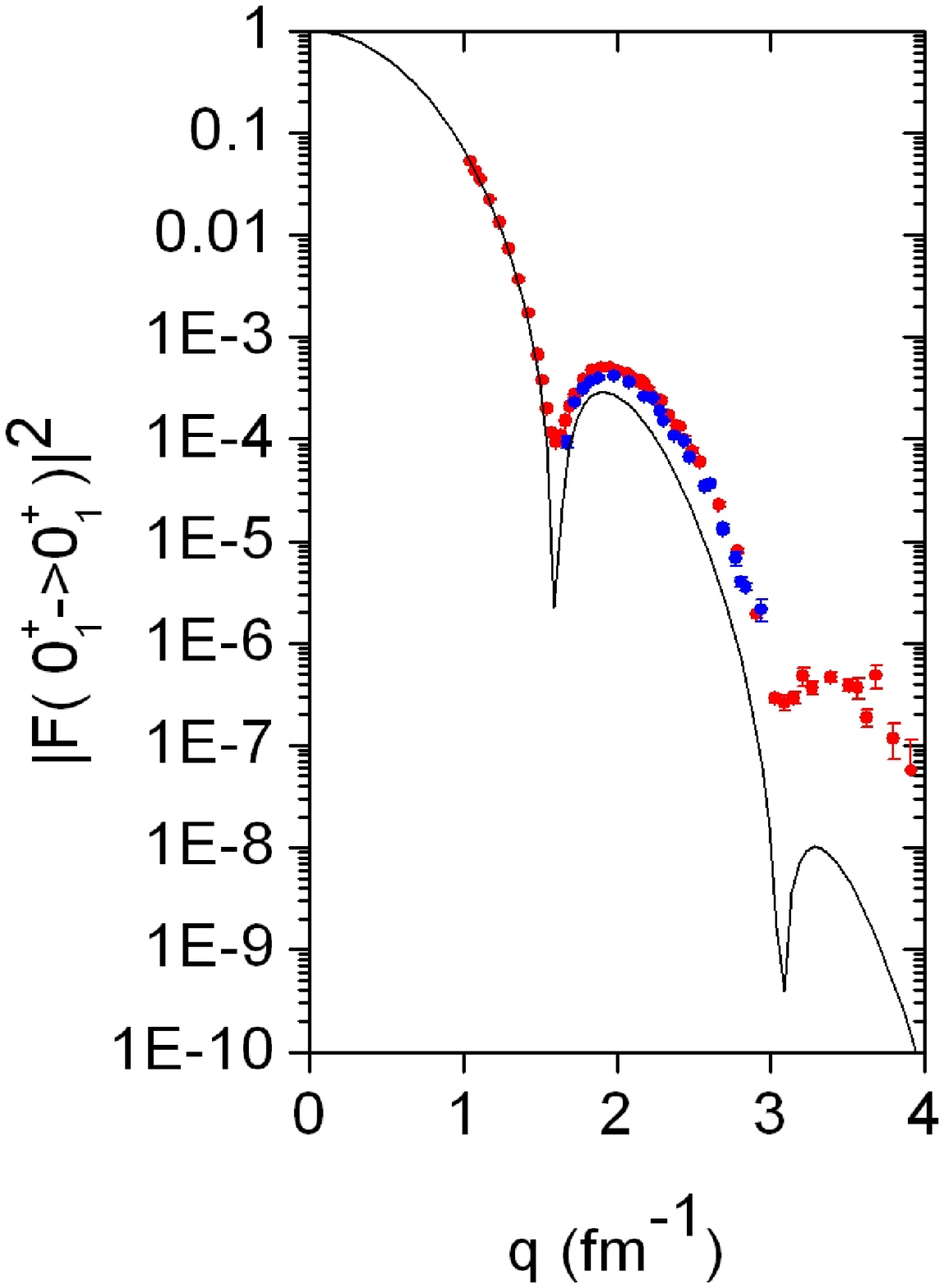,width=\linewidth}}
\end{minipage}\hfill
\begin{minipage}{.5\linewidth}
\centerline{\epsfig{file=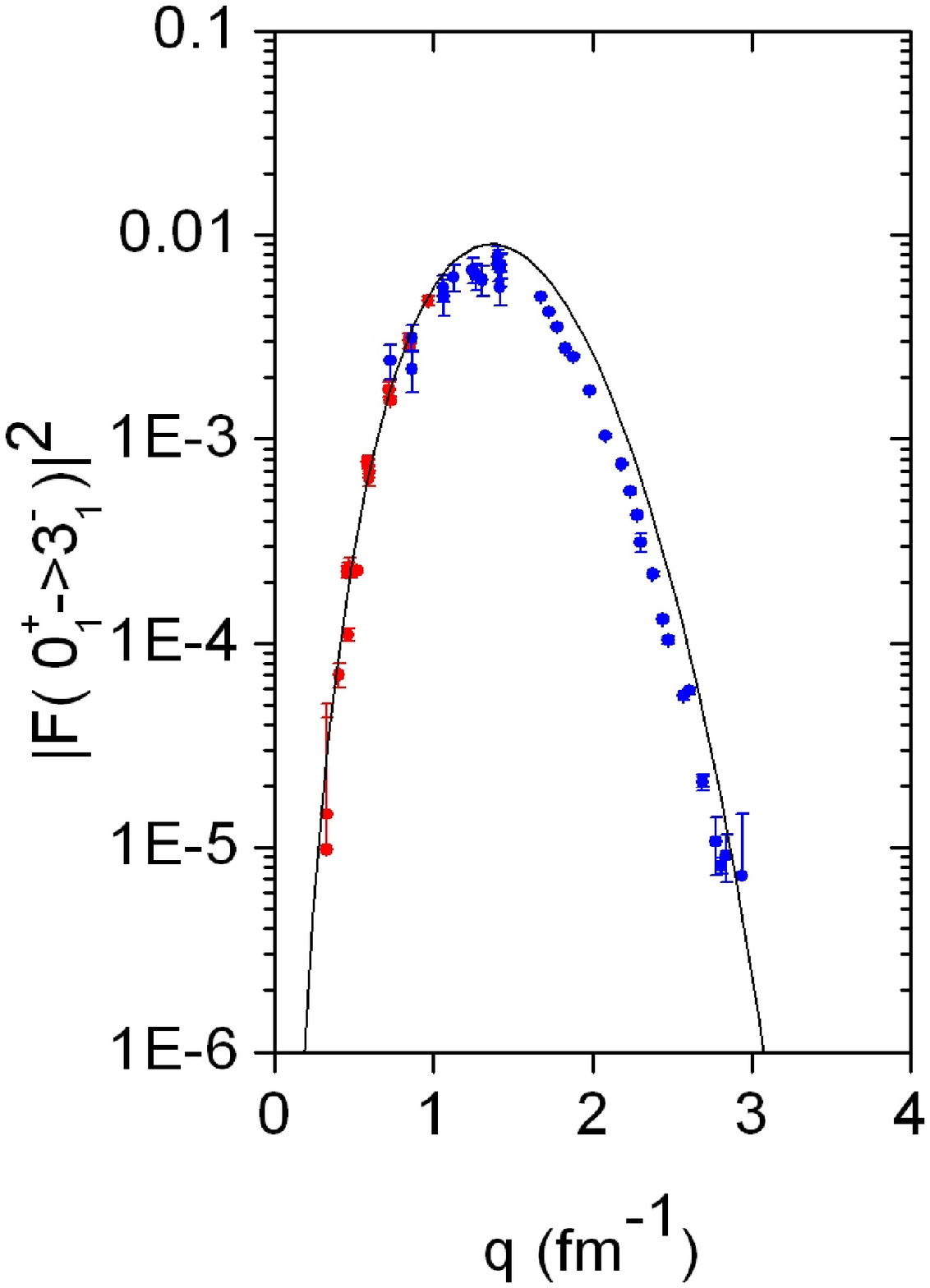,width=\linewidth}}
\end{minipage}
\vfill
\begin{minipage}{.5\linewidth}
\centerline{\epsfig{file=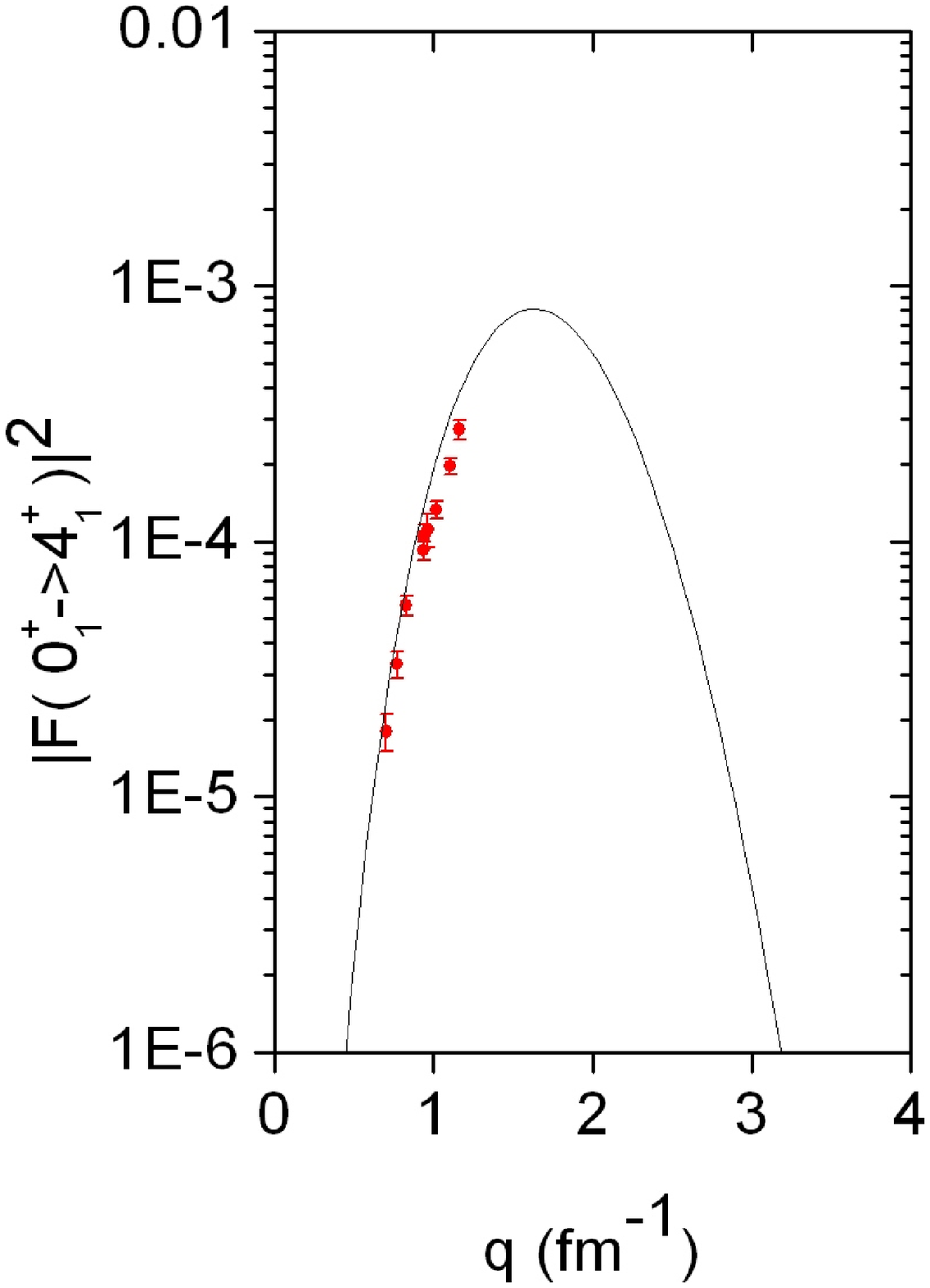,width=\linewidth}}
\end{minipage}\hfill
\begin{minipage}{.5\linewidth}
\centerline{\epsfig{file=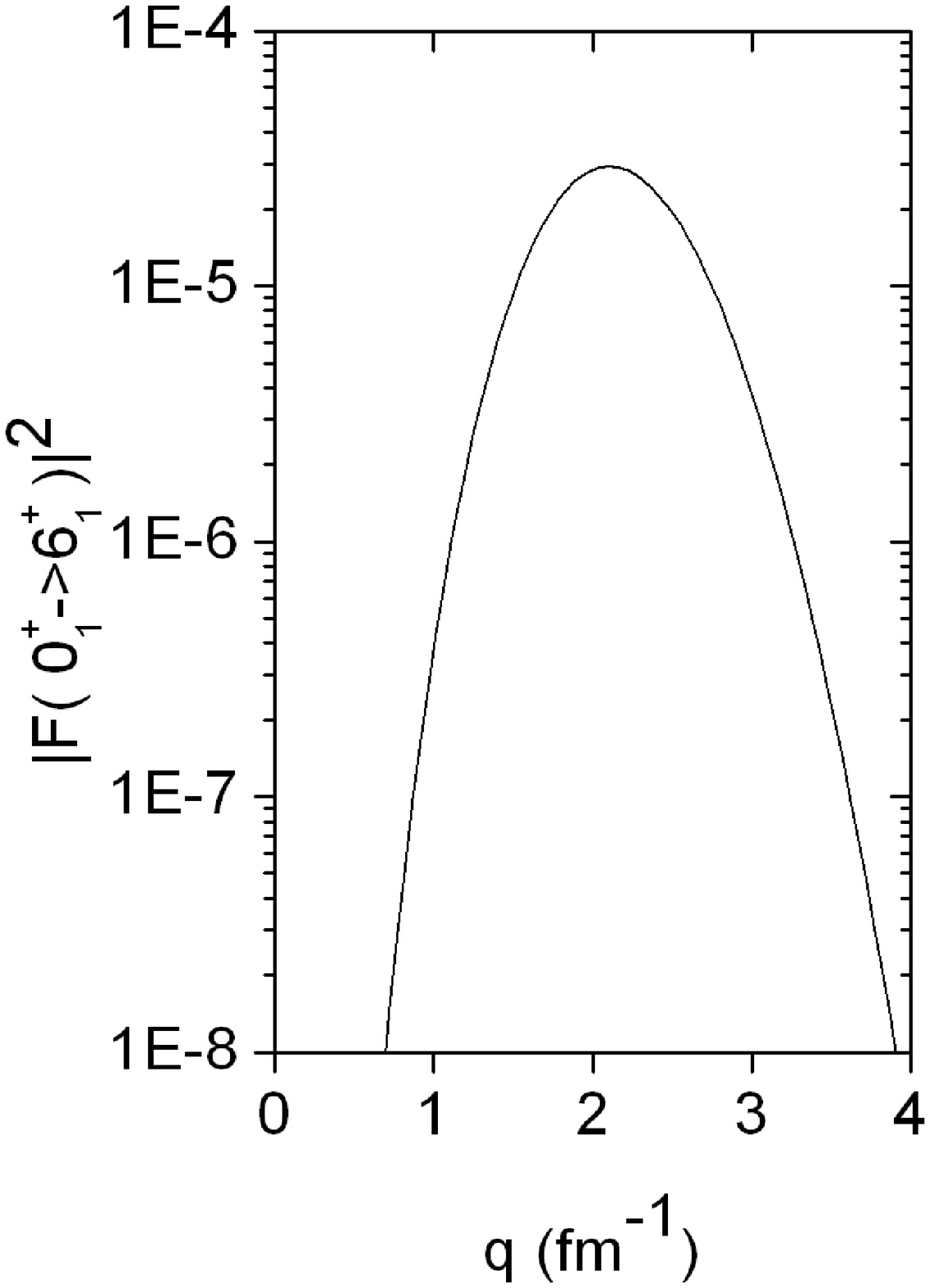,width=\linewidth}}
\end{minipage}
\caption[]{Comparison between the experimental form factors $|{\cal F}(0_1^+ \rightarrow L^P_i)|^2$ 
of $^{16}$O for the final states with $L^P_i=0^+_1$, $3^-_1$, $4^+_1$ and $6^+_1$ and those obtained 
for the spherical top with $N=10$ and $R^2=1.0$. The experimental data are taken from 
Refs.~\cite{Sick,Crannell,Bergstrom1,Bergstrom2,Stroetzel,Bishop}.}
\label{ffgsb}
\end{figure}

Form factors for electron scattering on $^{16}$O were measured long ago \cite{Sick}-\cite{Bishop}. 
For transitions along the ground state band the theoretical form factors for the spherical top 
are given in Table~\ref{fftop}. These expressions are only valid in the large $N$ limit. In this section, 
the form factors are calculated numerically for $N=10$ and $R^2=1$. In addition, the form factors are 
multiplied by an exponential factor $\exp(-q^2/4\alpha)$ for an extended distribution. The coefficient 
$\beta$ in Eq.~(\ref{trans}) is determined from the first minimum in the elastic form factor \cite{Sick} 
to be $\beta=2.071$ fm, and subsequently the coefficient $\alpha$ is determined from the charge radius 
of $^{16}$O \cite{Tilley} to be $\alpha=0.605$ fm$^{-2}$. 
If the $\alpha$-particle is not affected by the presence of the others, the coefficient $\alpha$ 
should be the same as for free $\alpha$-particles. The fitted value is slightly different from the 
free value $\alpha = 0.549$ fm$^{-2}$ \cite{helio1,helio2,helio3,helio4} indicating a polarization of 
the $\alpha$-particle in the medium. A comparison between experimental and calculated form factors 
in the ground state band is given in Fig.~\ref{ffgsb}. In the figure we have combined the experimental 
data for the excitation of the $3^-_1$ state (red) \cite{Bergstrom1,Stroetzel} with those of the unresolved 
doublet of the $3^-_1$ and $0^+_2$ states at 6.1 MeV (blue) \cite{Crannell,Bishop}. 

Electromagnetic transition rates along the ground state band can be obtained in the long wavelength 
limit of the form factors of Fig.~\ref{ffgsb} according to Eq.~(\ref{belif}). In Table~\ref{BELgsb}, 
the $B(EL)$ values are compared with experiment. The agreement is excellent and establishes the ground 
state band as a representation of the tetrahedral group ${\cal T}_d$ with $A_1$ symmetry. 
Finally, we note that the values presented in \cite{TdO16} are slightly different since those correspond 
to the analytic results in the large $N$ limit of Eq.~(\ref{BEL}) with $\beta = 2.0$ fm.

\begin{table}
\centering
\caption{Comparison of theoretical and experimental $B(EL)$ values in e$^2$fm$^{2L}$ and $E_{\gamma}$ 
values in keV, along the ground state band. The theoretical $B(EL)$ values are obtained in the long 
wavelength limit according to Eq.~(\ref{belif}). The $E_{\gamma}$ values are calculated from 
$E=0.511 \, L(L+1)$ MeV. The experimental values are taken from \cite{Tilley}.}
\vspace{5pt}
\label{BELgsb}
\begin{tabular}{cccccc}
\hline
\noalign{\smallskip}
$B(EL;L^P \rightarrow 0^+)$ & Th & Exp & $E_{\gamma}(L^P)$ & Th & Exp \\
\noalign{\smallskip}
\hline
\noalign{\smallskip}
$B(E3;3_1^- \rightarrow 0_1^+)$ &  215 & $205 \pm  11$ & $E_{\gamma}(3_1^-)$ &  6132 &  6130 \\
$B(E4;4_1^+ \rightarrow 0_1^+)$ &  425 & $378 \pm 133$ & $E_{\gamma}(4_1^+)$ & 10220 & 10356 \\
$B(E6;6_1^+ \rightarrow 0_1^+)$ & 9626 &               & $E_{\gamma}(6_1^+)$ & 21462 & 21052 \\ 
\noalign{\smallskip}
\hline
\noalign{\smallskip}
$\langle r^2 \rangle^{1/2}$ & $2.710$ & $2.710 \pm 0.015$ fm & & & \\ 
\noalign{\smallskip}
\hline
\end{tabular}
\end{table}

\begin{figure}
\centering
\begin{minipage}{.5\linewidth}
\centerline{\epsfig{file=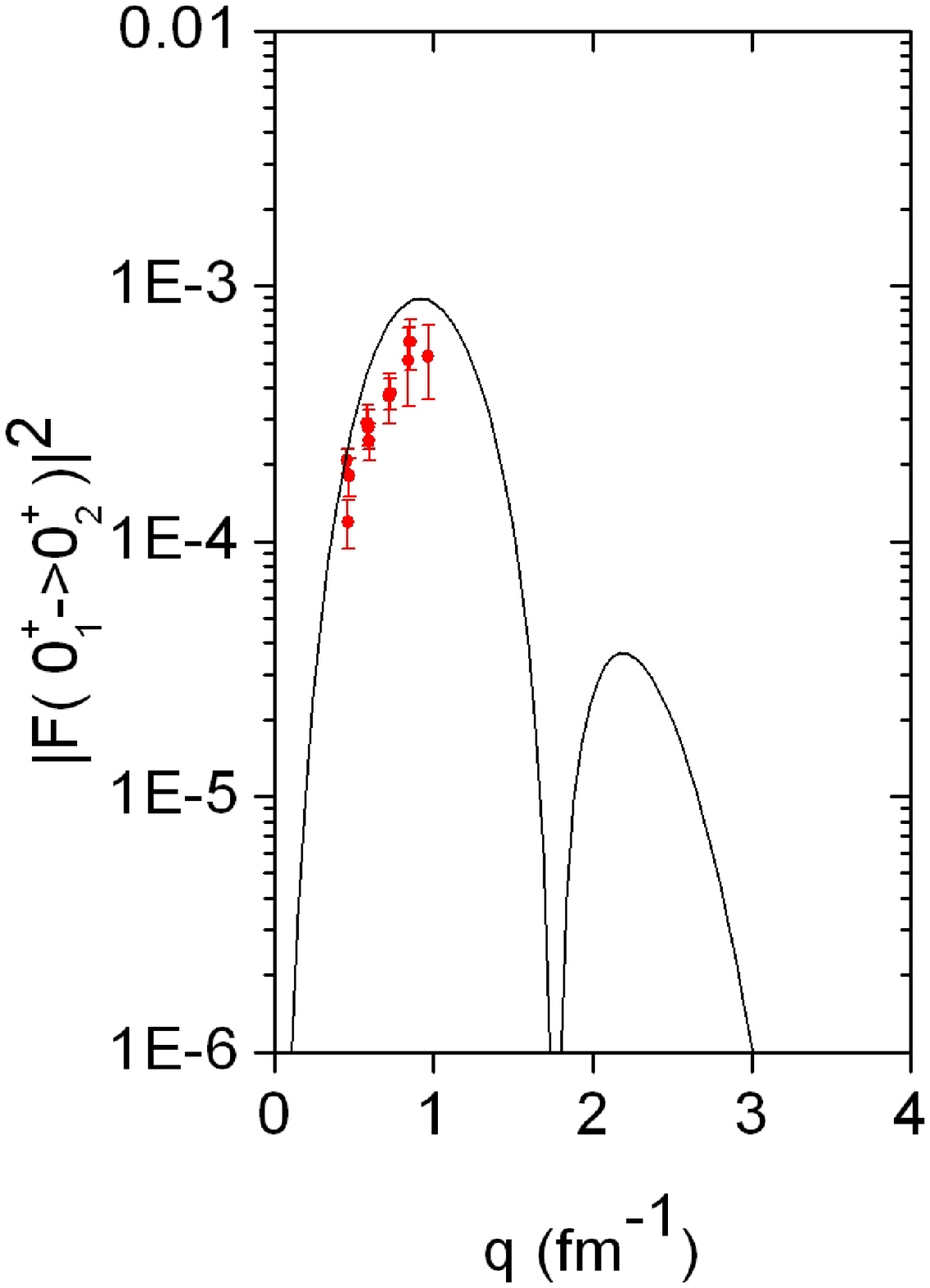,width=\linewidth}}
\end{minipage}\hfill
\begin{minipage}{.5\linewidth}
\centerline{\epsfig{file=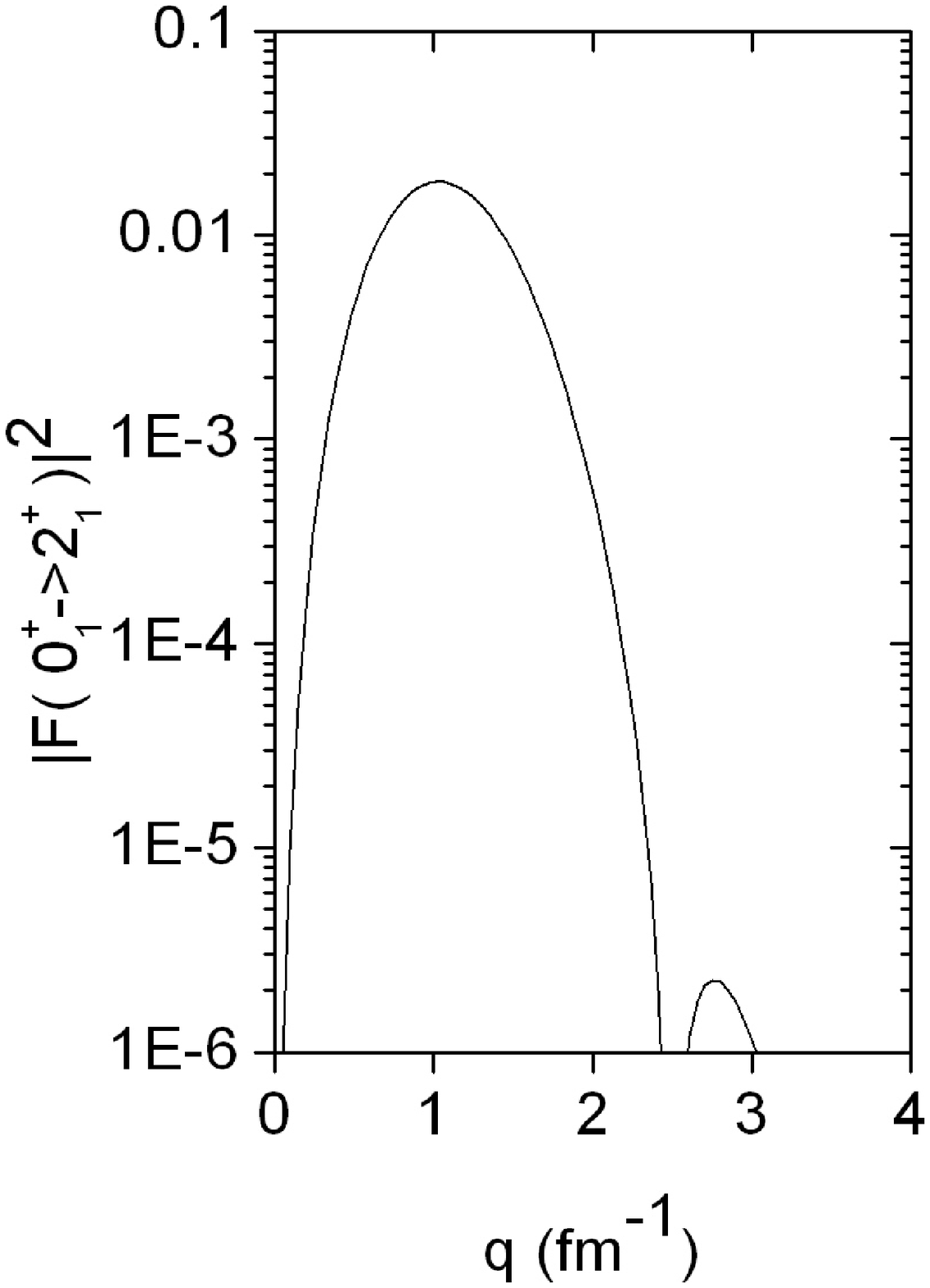,width=\linewidth}}
\end{minipage}
\vfill
\begin{minipage}{.5\linewidth}
\centerline{\epsfig{file=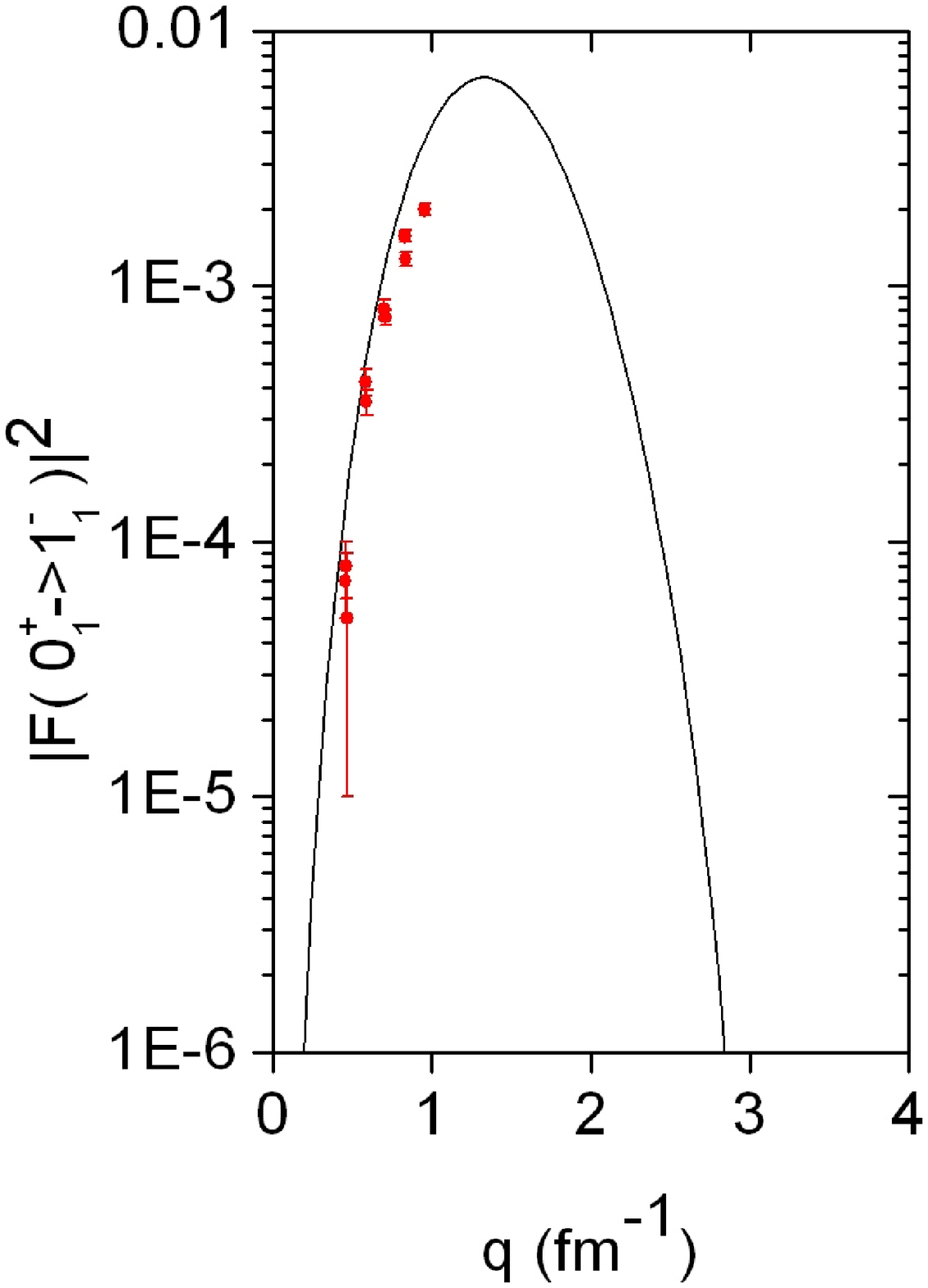,width=\linewidth}}
\end{minipage}
\caption[]{Comparison between the experimental form factors $|{\cal F}(0_1^+ \rightarrow L^P_i)|^2$ 
of $^{16}$O for the final states with $L^P_i=0^+_2$, $2^+_1$ and $1^-_1$ and those obtained 
for the spherical top with $N=10$ and $R^2=1.0$. 
The experimental data are taken from Ref.~\cite{Bergstrom1}.}
\label{ffvib}
\end{figure}

In the ACM, form factors and electromagnetic transition rates from the ground state to the vibrational 
states are reduced by a factor of $1/N$ with respect to the rotational excitations of the ground state band. 
In Fig.~\ref{ffvib} we show the results for transitions to the bandheads of the $A_1$, $E$ and $F_2$ vibrations. 

The corresponding electromagnetic transition rates can be obtained from the form factors in the long 
wavelength limit according to Eq.~(\ref{belif}). In Table~\ref{BELvib} they are compared with experiment for 
the same values of $R^2$. The $B(E1;1_1^- \rightarrow 0_1^+)$ value vanishes due to the tetrahedral symmetry: 
the initial and final states are symmetric $A_1$ whereas the dipole operator has symmetry $F_2$. 
The experimental value is rather small, in agreement with the ${\cal T}_d$ symmetry. The calculated 
monopole matrix element is a factor of 6 smaller than the experimental one, indicating that either 
the transition operator is insufficient to describe transitions from the ground state to the vibrations or 
that the nature of the vibrations is somewhat different from the assumed one. 
However, since the shape of the form factor is correctly given by the vibrational behavior and, as shown in 
Table~\ref{levels1}, one can identify a rotational band built on top of the state, we suggest 
that the assignment of the $0_2^+$ state as the bandhead of the $A_1$ vibration is correct, but that to 
obtain a better agreement one needs higher order terms in the transition operator \cite{work}. 

\begin{table}
\centering
\caption{Comparison between experimental and theoretical values for transitions between the ground state 
and states in the vibrational bands. The experimental values are taken from \cite{Tilley}.}
\label{BELvib}
\vspace{5pt}
\begin{tabular}{ccc}
\hline
\noalign{\smallskip}
$B(EL;L^P \rightarrow 0^+)$ & Th & Exp \\
\noalign{\smallskip}
\hline
\noalign{\smallskip}
$M(   0_1^+ \rightarrow 0_2^+)$ & 0.54 & $3.55 \pm 0.21$ fm$^2$ \\
$B(E1;1_1^- \rightarrow 0_1^+)$ & 0    & $(1.4 \pm 0.1) \times 10^{-4}$  e$^2$fm$^2$ \\ 
$B(E2;2_1^+ \rightarrow 0_1^+)$ & 26   & $7.4 \pm 0.2$ e$^2$fm$^4$ \\
\noalign{\smallskip}
\hline
\end{tabular}
\end{table}

Form factors and transition rates among excited states are in general more difficult to calculate.  
The $E2$ quadrupole transitions can be calculated numerically. The results are given in Table~\ref{BE2}. 
The agreement here is poor, especifically for the decays $2_1^+ \rightarrow 0_2^+$ and 
$4_1^+ \rightarrow 2_1^+$. With the assignment of $4_1^+$ as a member of the ground state rotational band, 
the $B(E2;4_1^+ \rightarrow 2_1^+)$ value vanishes due to a selection rule for the index spin. 
All members of the ground state band have $I=L$, so the index spin of the initial state is $I=4$, whereas 
the $2_1^+$ state has $I=0$. Since the transition operator $\hat D$ is a vector in index spin, the 
$B(E2)$ value vanishes. The experimental value is large. 
On the other hand, the calculated $B(E2;4_2^+ \rightarrow 2_1^+)$ value is large. Therefore, 
either the assignment of $4_1^+$ and $4_2^+$ should be interchanged, in which case an attempt 
should be made to measure the $B(E4;4_2^+ \rightarrow 0_1^+)$, or the two states are strongly 
mixed due to their vicinity in energy as seen in Fig.~\ref{O16}.

The same situation occurs for the $2_1^+$ and $2_2^+$ states which we have assigned to $(010)E$ and 
$(001)F_2$. Therefore, while the assignments of the bandheads of the vibrational bands may be correct, 
it is not clear whether the states built on them are correctly assigned, a situation similar to that 
encountered in $^{12}$C \cite{ACM2}.

\begin{table}
\centering
\caption{Comparison between experimental and theoretical $B(E2)$ values in e$^2$fm$^4$.  
The experimental values are taken from \cite{Tilley}.}
\label{BE2}
\vspace{5pt}
\begin{tabular}{ccc}
\hline
\noalign{\smallskip}
$B(EL;L^P \rightarrow L'^{P'})$ & Th & Exp \\
\noalign{\smallskip}
\hline
\noalign{\smallskip}
$B(E2;2_1^+ \rightarrow 0_2^+)$ &  6 & $ 65   \pm  7$ \\ 
$B(E2;4_1^+ \rightarrow 2_1^+)$ &  0 & $156   \pm 14$ \\ 
$B(E2;4_2^+ \rightarrow 2_1^+)$ & 36 & $  2.4 \pm  0.7$ \\ 
$B(E2;1_1^- \rightarrow 3_1^-)$ & 19 & $ 50.3 \pm 12.0$ \\ 
$B(E2;2_1^- \rightarrow 3_1^-)$ & 10 & $ 19.6 \pm  1.7$ \\ 
$B(E2;2_1^- \rightarrow 1_1^-)$ &  8 & $ 24.7 \pm  3.6$ \\ 
\noalign{\smallskip}
\hline
\end{tabular}
\end{table}

\section{Summary and conclusions}

In this paper, we have introduced an algebraic description of the four-body problem 
in terms of the spectrum generating algebra (SGA) of $U(10)$ based on the bosonic 
representation of the Jacobi vectors, $\vec{\rho}$, $\vec{\lambda}$ and $\vec{\eta}$. 
In particular, we have shown that the ACM for four-body clusters contains the spherical 
top with tetrahedral ${\cal T}_d$ symmetry as a special solution. The spherical top has 
been applied to the study of the spectrum, form factors and electromagnetic transition 
rates of the nuclear $^{16}$O, as composed of four $\alpha$-particles with tetrahedral 
symmetry. Evidence for this symmetry is particularly strong for the ground state band 
$(000)A_1$ with a rotational sequence $L^P=0^+$, $3^-$, $4^+$, $6^+$, and weaker for the 
excited bands $(100)A_1$, $(010)E$ and $(001)F_2$. 

\begin{figure}
\centering
\vspace{15pt}
\setlength{\unitlength}{0.7pt}
\begin{picture}(460,220)(0,0)
\thicklines
\put(110, 50) {\circle*{10}}
\put(150, 50) {\circle*{10}}
\put(190, 50) {\circle*{10}}
\put(230, 50) {\circle*{10}}
\put(110, 50) {\line( 1,0){120}}
\put(150,  0) {Linear: \bf ${\cal C}_{\infty v}$}
\put(310, 30) {\circle*{10}}
\put(350, 30) {\circle*{10}}
\put(310, 70) {\circle*{10}}
\put(350, 70) {\circle*{10}}
\put(310, 30) {\line( 1,0){40}}
\put(310, 70) {\line( 1,0){40}}
\put(310, 30) {\line( 0,1){40}}
\put(350, 30) {\line( 0,1){40}}
\put(300,  0) {Square: \bf ${\cal D}_{4h}$}
\put( 60,150) {\circle*{10}} 
\put(120,150) {\circle*{10}}
\put(135,180) {\circle*{10}}
\put( 90,210) {\circle*{10}}
\put( 60,150) {\line( 1,0){60}}
\put( 60,150) {\line( 1,2){30}}
\put(120,150) {\line(-1,2){30}}
\put(120,150) {\line( 1,2){15}}
\put(135,180) {\line(-3,2){45}}
\multiput( 60,150)(5,2){16}{\circle*{2}}
\put( 60,120) {Tetrahedral: \bf ${\cal T}_d$}
\put(200,150) {\circle*{10}} 
\put(260,150) {\circle*{10}}
\put(275,180) {\circle*{10}}
\put(230,195) {\circle*{10}}
\put(200,150) {\line( 1,0){60}}
\put(200,150) {\line( 2,3){30}}
\put(260,150) {\line(-2,3){30}}
\put(260,150) {\line( 1,2){15}}
\put(275,180) {\line(-3,1){45}}
\multiput(200,150)(5,2){16}{\circle*{2}}
\put(200,120) {Pyramidal: \bf ${\cal C}_{3v}$}
\put(340,150) {\circle*{10}} 
\put(400,150) {\circle*{10}}
\put(370,170) {\circle*{10}}
\put(370,210) {\circle*{10}}
\put(340,150) {\line( 3,2){30}}
\put(400,150) {\line(-3,2){30}}
\put(370,170) {\line( 0,1){40}}
\put(340,120) {Planar: \bf ${\cal D}_{3h}$}
\end{picture}
\vspace{15pt}
\caption[]{Four-body configurations: tetrahedral, pyramidal, planar, linear and square 
with their respective point-group symmetries.}
\label{shapes}
\end{figure}
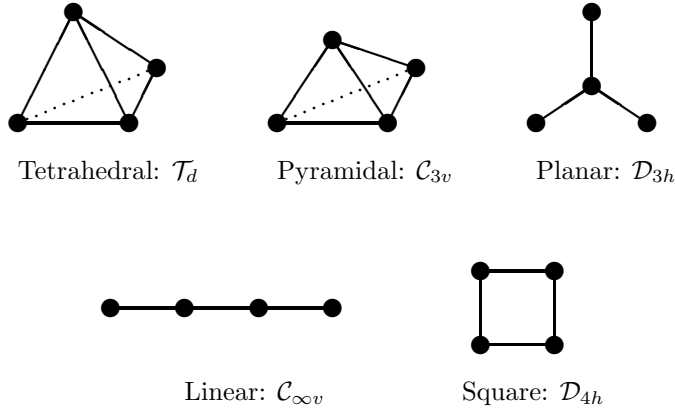

The SGA of $U(10)$ can be used for any four-body problem whether the particles are 
identical or not, and for any discrete symmetry, be it ${\cal T}_d$ or other, 
and therefore it relevant for applications to other four-body problems in molecular, 
nuclear and hadronic physics, both for rigid and non-rigid configurations. 
Particularly simple are the descriptions of the rigid configurations of Fig.~\ref{shapes}, 
tetrahedral, pyramidal and planar, and those of the harmonic oscillator ($U(9)$ limit) and 
the deformed oscillator ($SO(10)$ limit). Other configurations, such as the linear chain and 
the square are somewhat more involved, but they can be dealt with as well.

Finally, a variation of the algebraic method in which the Jacobi variables $\vec{\rho}$, 
$\vec{\lambda}$ and $\vec{\eta}$ are not fixed to their equilibrium values, as in the rigid 
configuration of Fig.~\ref{shapes}, but are allowed to perform large amplitude motion can be used
to describe quasi-molecular configurations. For $\alpha + ^{12}$C, the variables $\vec{\rho}$ 
and $\vec{\lambda}$ of Fig.~\ref{shape} need to be kept rigid as in our previous paper on $^{12}$C 
\cite{ACM2}, while the variable $\vec{\eta}$ may undergo large amplitude motion. A suitable 
algebraic description for these configurations is $U(7) \otimes U(4)$, where $U(7)$ describes 
the structure of $^{12}$C and $U(4)$ the relative motion between $\alpha$ and $^{12}$C \cite{Iachello}.  

\section*{Acknowledgments}

This work was supported in part by research grant IN107314 from PAPIIT-DGAPA and 
in part by DOE Grant DE-FG02-91ER40608.

\end{document}